\documentclass[aip,reprint,floatfix]{revtex4-1}
\usepackage[utf8]{inputenc}
\usepackage{amssymb}
\usepackage{amsthm}
\usepackage{amsmath}
\usepackage{graphicx}
\usepackage{hyperref}
\usepackage{bm}
\usepackage{xspace}

\newcommand{\mvec}[1]{\bm{#1}}
\newcommand{\bnabla}{{\bm{\nabla}}}
\newcommand{\D}{\mathrm{d}}

\makeatletter
\g@addto@macro{\UrlBreaks}{%
\do\/\do\a\do\b\do\c\do\d\do\e\do\f%
\do\g\do\h\do\i\do\j\do\k\do\l\do\m%
\do\n\do\o\do\p\do\q\do\r\do\s\do\t%
\do\u\do\v\do\w\do\x\do\y\do\z%
\do\A\do\B\do\C\do\D\do\E\do\F\do\G%
\do\H\do\I\do\J\do\K\do\L\do\M\do\N%
\do\O\do\P\do\Q\do\R\do\S\do\T\do\U%
\do\V\do\W\do\X\do\Y\do\Z}
\makeatother

\newcommand\esl{ESL\xspace}
\newcommand\eslbundle{ESL Bundle\xspace}
\newcommand\esldemo{ESL Demonstrator\xspace}

\newcommand\psolver{PSolver\xspace}
\newcommand\flook{flook\xspace}
\newcommand\elsi{ELSI\xspace}
\newcommand\pspio{pspio\xspace}
\newcommand\elpa{ELPA\xspace}
\newcommand\pexsi{PEXSI\xspace}
\newcommand\ntpoly{NTPoly\xspace}
\newcommand\sips{SLEPc-SIPs\xspace}
\newcommand\libpsml{libPSML\xspace}
\newcommand\libfdf{LibFDF\xspace}
\newcommand\libxc{Libxc\xspace}
\newcommand\libvdwxc{libvdwxc\xspace}
\newcommand\libgridxc{libGridXC\xspace}
\newcommand\libomm{LibOMM\xspace}
\newcommand\matrixswitch{MatrixSwitch\xspace}
\newcommand\escdf{ESCDF\xspace}
\newcommand\libescdf{libescdf\xspace}
\newcommand\wannier{\texttt{wannier90}\xspace}

\newcommand\siesta{\textsc{Siesta}\xspace}
\newcommand\fhiaims{FHI-aims\xspace}
\newcommand\octopus{Octopus\xspace}
\newcommand\abinit{\textsc{ABINIT}\xspace}
\newcommand\bigdft{BigDFT\xspace}
\newcommand\gpaw{\textsc{GPAW}\xspace}

\begin{document}

\title{The CECAM Electronic Structure Library and the modular software development paradigm}

\author{Micael J.~T.~Oliveira}
\email{micael.oliveira@mpsd.mpg.de}
\affiliation{Max Planck Institute for the Structure and Dynamics of Matter,
             D-22761 Hamburg, Germany}

\author{Nick Papior}
\email{nickpapior@gmail.com}
\affiliation{DTU Computing Center, Technical University of Denmark, 
             2800 Kgs. Lyngby, Denmark}

\author{Yann Pouillon}
\email{yann.pouillon@materialsevolution.es}
\affiliation{Departamento CITIMAC, Universidad de Cantabria, Santander, Spain} 
\affiliation{Simune Atomistics, 20018 San Sebasti\'an, Spain}

\author{Volker Blum} 
\affiliation{Department of Mechanical Engineering and Materials Science, 
             Duke University, Durham, NC 27708, USA}
\affiliation{Department of Chemistry, 
             Duke University, Durham, NC 27708, USA}

\author{Emilio Artacho} 
\affiliation{CIC Nanogune BRTA and DIPC, 20018 San Sebasti\'an, Spain}
\affiliation{Ikerbasque, Basque Foundation for Science, 48011 Bilbao, Spain}
\affiliation{Theory of Condensed Matter,
             Cavendish Laboratory, University of Cambridge, 
             Cambridge CB3 0HE, United Kingdom}
             
\author{Damien Caliste} 
\affiliation{Department of Physics, IRIG, Univ. Grenoble Alpes and CEA, 
             F-38000 Grenoble, France.}

\author{Fabiano Corsetti} 
\affiliation{Departments of Materials and Physics, and the Thomas Young 
             Centre for Theory and Simulation of Materials, 
             Imperial College London, London SW7 2AZ, United Kingdom}
\affiliation{Synopsys Denmark, 2100 Copenhagen, Denmark}

\author{Stefano de Gironcoli} 
\affiliation{Scuola Internazionale Superiore di Studi Avanzati,
             34136 Trieste, Italy}

\author{Alin M. Elena} 
\affiliation{Scientific Computing Department, Daresbury Laboratory,
             Warrington WA4 4AD, United Kingdom}

\author{Alberto Garc\'ia} 
\affiliation{Institut de Ci\`encia de Materials de Barcelona (ICMAB-CSIC),
             Bellaterra E-08193, Spain}

\author{V\'{\i}ctor M. Garc\'ia-Su\'arez} 
\affiliation{Departamento de F\'isica, Universidad de Oviedo \& CINN, 
             33007 Oviedo, Spain}

\author{Luigi Genovese} 
\affiliation{Department of Physics, IRIG, Univ. Grenoble Alpes and CEA, 
             F-38000 Grenoble, France.}

\author{William P. Huhn} 
\affiliation{Department of Mechanical Engineering and Materials Science, 
             Duke University, Durham, NC 27708, USA}

\author{Georg Huhs} 
\affiliation{Barcelona Supercomputing Center (BSC), 08034 Barcelona, Spain}

\author{Sebastian Kokott} 
\affiliation{Fritz Haber Institut, 14195 Berlin, Germany}

\author{Emine K\"{u}\c{c}\"{u}kbenli} 
\affiliation{Scuola Internazionale Superiore di Studi Avanzati,
             34136 Trieste, Italy}
\affiliation{John A. Paulson School of Engineering and Applied Sciences,
             Harvard University, Cambridge, Massachusetts 02138, USA}

\author{Ask H.~Larsen} 
\affiliation{Nano-Bio Spectroscopy Group and ETSF, Departamento de 
             F\'isica de Materiales, Universidad del Pa\'is Vasco UPV/EHU, 
             20018 San Sebasti\'an, Spain}
\affiliation{Simune Atomistics, 20018 San Sebasti\'an, Spain}

\author{Alfio Lazzaro} 
\affiliation{Department of Chemistry, University of Z\"urich,
             CH-8057 Z\"urich, Switzerland}

\author{Irina V. Lebedeva} 
\affiliation{CIC Nanogune BRTA, 20018 San Sebasti\'an, Spain}

\author{Yingzhou Li} 
\affiliation{Department of Mathematics, Duke University, 
             Durham, NC 27708-0320, USA}

\author{David L\'opez-Dur\'an} 
\affiliation{CIC Nanogune BRTA, 20018 San Sebasti\'an, Spain}

\author{Pablo L\'opez-Tarifa} 
\affiliation{Centro de F\'isica de Materiales, Centro Mixto CSIC-UPV/EHU,
             20018 San Sebasti\'an, Spain}

\author{Martin L\"uders} 
\affiliation{Max Planck Institute for the Structure and Dynamics of Matter, 
             D-22761 Hamburg, Germany}
\affiliation{Scientific Computing Department, Daresbury Laboratory,
             Warrington WA4 4AD, United Kingdom}

\author{Miguel A.~L.~Marques} 
\affiliation{Institut f\"ur Physik, Martin-Luther-Universit\"at
             Halle-Wittenberg, 06120 Halle (Saale), Germany}

\author{Jan Minar} 
\affiliation{New Technologies Research Centre, University of West Bohemia, 
             301 00 Plzen, Czech Republic}

\author{Stephan Mohr} 
\affiliation{Barcelona Supercomputing Center (BSC), 08034 Barcelona, Spain}

\author{Arash A. Mostofi} 
\affiliation{Departments of Materials and Physics, and the Thomas Young 
             Centre for Theory and Simulation of Materials, 
             Imperial College London, London SW7 2AZ, United Kingdom}

\author{Alan O'Cais} 
\affiliation{Institute for Advanced Simulation (IAS), 
             J\"ulich Supercomputing Centre (JSC), Forschungszentrum 
             J\"ulich GmbH, 52425 J\"ulich, Germany}

\author{Mike C. Payne} 
\affiliation{Theory of Condensed Matter,
             Cavendish Laboratory, University of Cambridge, 
             Cambridge CB3 0HE, United Kingdom}
             
\author{Thomas Ruh} 
\affiliation{Institute of Materials Chemistry, TU Wien, 1060 Vienna, Austria}
             
\author{Daniel G. A. Smith} 
\affiliation{Molecular Sciences Software Institute,
             Blacksburg, Virginia 24060, USA}

\author{Jos\'e M.~Soler} 
\affiliation{Departamento e Instituto de F\'isica de la Materia Condensada
             (IFIMAC), Universidad Aut\'onoma de Madrid, 28049 Madrid, Spain}

\author{David A. Strubbe} 
\affiliation{Department of Physics, University of California,
             Merced, CA 95343, USA} 

\author{Nicolas Tancogne-Dejean} 
\affiliation{Max Planck Institute for the Structure and Dynamics of Matter, 
             D-22761 Hamburg, Germany}
             
\author{Dominic Tildesley} 
\affiliation{School of Chemistry, University of Southampton,
             Southampton, SO17 1BJ, United Kingdom}

\author{Marc Torrent} 
\affiliation{CEA, DAM, DIF, F-91297 Arpajon, France}
\affiliation{Universit\'e Paris-Saclay, CEA, Laboratoire Mati\`ere en Conditions Extr\^emes, 
             91680 Bruy\`eres-le-Ch\^atel, France}

\author{Victor Wen-zhe Yu} 
\affiliation{Department of Mechanical Engineering and Materials Science, 
             Duke University, Durham, NC 27708, USA}

\date{6 June 2020, accepted by J. Chem. Phys. 8 June 2020, to appear in
https://doi.org/10.1063/5.0012901}

\begin{abstract}
  First-principles electronic structure calculations 
are now accessible to a very large community of users across many 
disciplines thanks to many successful software packages,
some of which are described in this special issue.
  The traditional coding paradigm for such packages is \emph{monolithic},
i.e., regardless of how modular its internal structure may be, the code is built 
independently from others, essentially from the compiler up, 
possibly with the exception of linear-algebra and message-passing libraries.
  This model has endured and been quite successful for decades. The successful 
evolution of the electronic structure methodology itself, however, has resulted 
in an increasing complexity and an ever longer list of features expected within all 
software packages, which implies a growing amount of replication between 
different packages, not only in the initial coding but, more importantly, 
every time a code needs to be re-engineered to adapt to the evolution 
of computer hardware architecture.
  The Electronic Structure Library (\esl) was initiated by CECAM 
(the European Centre for Atomic and Molecular Calculations) 
to catalyze a paradigm shift away from the monolithic model and promote 
modularization, with the ambition to extract common tasks from 
electronic structure codes and redesign them as open-source libraries 
available to everybody. 
  Such libraries include, e.g., ``heavy-duty'' ones that have the potential for 
a high degree of parallelisation
and adaptation to novel hardware \textit{within them}, thereby separating the
sophisticated computer science aspects of performance optimization and 
re-engineering from the computational science done by, e.g., physicists 
and chemists when implementing new ideas.
  We envisage that this modular paradigm will improve overall coding 
efficiency and enable specialists (whether they be computer scientists 
or computational scientists) to use their skills more effectively, and 
will lead to a more dynamic evolution of software in the community as 
well as lower barriers to entry for new developers.
  The model comes with new challenges, though. The building and compilation of a 
code based on many interdependent libraries (and their versions) is a much 
more complex task than that of a code delivered in a single self-contained package. 
  Here we describe the state of the \esl, the different libraries it now contains, 
the short- and mid-term plans for further libraries, and the way the new challenges
are faced. 
  The \esl is a community initiative into which several pre-existing codes 
and their developers have contributed with their software and efforts, 
from which several codes are already benefiting, and which remains open 
to the community.
\end{abstract}

\maketitle

\section{Introduction}
\label{sec:intro}

Electronic structure theory is among the most productive branches of computational science today.~\cite{Mavropoulos2017} The necessary underlying level of theory -- Dirac's Equation -- is analytically known exactly.~\cite{Dirac1929} It is applicable to condensed matter physics, chemistry, materials science and, in fact, touches all branches of engineering -- whenever either modified or completely new technologically more capable materials are needed. Practical, i.e., numerically tractable, approximations to Dirac's Equation can be used to predict the properties of molecules, solids, liquids, interfaces, including their responses to environmental stimuli (fields, currents, mechanical stimuli, etc.). They typically provide sufficient accuracy and reliability~\cite{Mardirossian2017} to formulate experimentally testable hypotheses and, ultimately, accelerate the discovery and development of ``new'' molecules and materials. 
The growth of the field is reflected in a plethora of existing and new software developments that implement aspects of electronic structure theory either for specialized or rather broad general use cases. The community-wide \texttt{psi-k.net} website lists over thirty ``codes'' at the time of writing (December 2019) and 74 individual code projects are listed at the ``Community Code Database'' of the U.S. based Molecular Software Sciences Institute (MolSSI),~\cite{molssi} another community-bridging organization working to support a broad set of ``codes'' and their users.~\cite{Wilkins-Diehr2018,Krylov2018}

While the electronic structure community (ESC) is thus extremely active in developing software that enables a host of scientific insights, developments have historically occurred in the form of different individual software packages that are largely distinct from each another at the code level. A notable exception are numerical and/or performance related libraries which are often generic to the broader computational community, e.g., basic linear algebra subroutines (BLAS),\cite{blas} exploiting parallelism at the message passing interface (MPI) level,\cite{mpi} higher-level linear algebra utilities (most importantly LAPACK\cite{lapack_anderson_1999} and its parallel counterpart, ScaLAPACK\cite{scalapack}) or fast Fourier transforms (FFTW).\cite{FFTW05} 

\begin{figure}[t]
\includegraphics[width=0.99\linewidth]{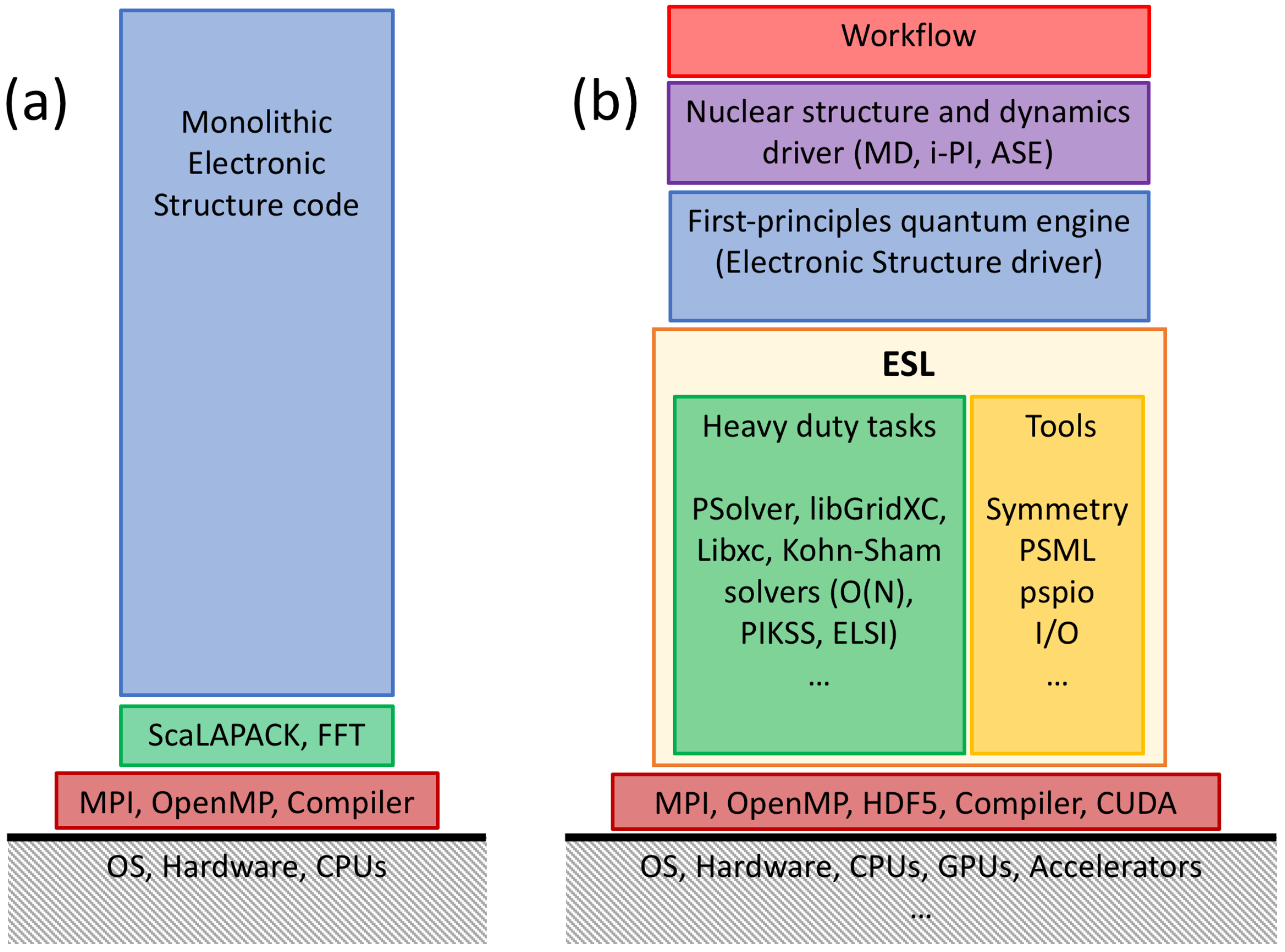}
\caption{Comparison of the traditional 
monolithic and the emerging modular paradigms in electronic structure
coding.
  One of today's electronic-structure codes  
-- large blue box in panel (a) -- thins down 
into the higher-level electronic structure driver that
defines the particular code -- blue box in 
panel (b), allowing for a more specialized and sustainable 
development of the different parts of the software.
  The acronyms in the figure indicate operating system (OS),
fast Fourier transforms (FFT), input/output (I/O), molecular
dynamics (MD), and linear scaling with the number of atoms, 
$\mathcal O(N)$, respectively, in addition to the central and graphical 
processing units, CPU and GPU, respectively.
  Steering upper-level drivers are distinguished between 
versatile toolkits such as ASE, and handlers of massive 
amounts of quantum-engine replicas, such as i-PI.
  In addition to the well-known MPI, OpenMP, CUDA, and HDF
libraries, other acronyms and names relate to libraries 
described in Sections~\ref{sec:intro}, \ref{sec:shared}, 
and \ref{sec:implement}.
\label{fig:models}} 
\end{figure}

ESC software development has historically taken place within a model of largely monolithic programs, in which, on top of a main quantum engine,
all further developments are incorporated incrementally.
  The (now) more traditional electronic structure codes are steadily growing, each incorporating all or many of
the developments that have become standard in the community. 
  This model is illustrated in Fig.~\ref{fig:models}(a). 
  Furthermore, each code needs re-engineering to adapt to the constant hardware evolution, most notably in 
high-performance computing, and most of the re-engineering is carried out on tasks that are
common to all or most of the codes.
  In addition to this obvious inefficiency, two other important problems are inherent in 
the monolithic model.
  Firstly, it stifles innovation: novel methodological (physics) ideas within the wider community 
can only be implemented by joining any of the pre-existing efforts. It is increasingly hard to
start a project from scratch. 
  This problem is partly addressed by the open-source model of programming, well established in some modern
electronic structure projects, to which novel ideas can be incorporated by external coders, 
at least in principle (note, however, that poor quality, undocumented open source code does not fulfill this requirement).
  Secondly, the monolithic model allows very little differentiation in the profiles
of human resources needed for the project: there is a need for people with expertise in the state-of-the-art
for both computational science (e.g., physics, chemistry, etc.) and computer science 
(e.g., software engineering, performance optimisation, hardware architecture, numerical analysis etc.). 

\section{Shared libraries and the \esl}
\label{sec:shared}

\subsection{The library sharing movement}

  Partly in response to the problems mentioned above, and partly 
following the spirit of the open access movement and inspired by
well established practices in software engineering, 
the computational physics and chemistry communities 
have witnessed the appearance of libraries---understanding this term broadly---which perform particular, well-defined tasks that are 
common to many codes.
  We will not review this movement here, but will illustrate it with some examples from the ESC.
  Take, for instance, the exploitation of symmetry in computational simulation of 
both molecular and crystalline systems. This involves a well-defined set of tasks, 
from recognising the symmetry group for a specified structure, to the labelling 
of eigenstates according to irreducible representations, including the reduction 
of the eigenproblem complexity, or the optimisation of Brillouin-zone sampling.
  Several libraries have appeared within this free sharing movement to perform
these tasks (e.g., the \texttt{spglib} library~\cite{togo2018}).

  The handling of symmetry is an example of a very general
pre- and post-processing tool whose function can be defined 
completely independently and is one of many other similar possibilities 
that constitute opportunities for creating libraries.
  Another notable case is that of \wannier,~\cite{wannier-website} which not only calculates
maximally localised Wannier functions~\cite{MV_PRB56} from the outputs of 
electronic structure codes, but can also determine many properties using these Wannier functions.
  Remarkably, the authors' ambition from the very beginning of the project was
to maximise its applicability to all classes of electronic structure methods, and they have
managed to limit code dependencies to an absolute minimum. This has enabled a
very widespread adoption within the wider electronic structure community (see Sec.~\ref{sec:wannier}).

  In addition to these kinds of tool, other sharing/library efforts have 
been appearing which we can characterise as top-level steering codes, and
low-level routines, as shown in Fig.~\ref{fig:models}(b),
which illustrates the new emerging paradigm. 
  Among the former is the integration of electronic structure codes as 
``solvers'' or ``quantum engines" into broader frameworks, typically 
handling the nuclear degrees of freedom, such as the python-based
Atomic Simulation Environment (ASE)~\cite{ase-paper} or the i-PI
framework~\cite{KAPIL2019214} for classical and path-integral molecular dynamics.
Both of these support a large number of underlying electronic structure codes.
Also in the top-level category, much effort is now being dedicated to
general-purpose workflow tools that steer and automate the running of
electronic structure codes in complex procedures and encompass the ambition of 
versatility (see, e.g., the AiiDA project~\cite{PIZZI2016218}), providing a 
much more detailed picture of how electronic structure methods are applied 
as compared to a decade ago.
  As a pioneering example of a low-level shared library, we mention
\libxc~\cite{Marques2012_2272,Lehtola2018_1}, which implements 
hundreds of local and semilocal
exchange-correlation functionals and is now very widely used (see Sec.~\ref{sec:libxc}).
  
  There are many other tasks and needs in electronic structure that 
may be generically abstracted in the form of shared libraries, 
with common frameworks and shared workloads in order to more readily 
achieve maturity of established functionality, numerical correctness, 
and continued development of new functionality at the same time. 
  Developing electronic-structure software based on common standards, 
libraries, application programming interfaces (APIs), and flexible software components is a trend that is 
therefore gaining prominence in the field. 

Additionally, at a social level, such shared developments bring different communities together and reinforce existing collaborations within the communities themselves. Significant challenges on this path are often simple, related to human time and workload and include: identifying and locating an existing solution to a code problem at the time when it is needed; finding and reading documentation to understand and co-develop software originally written by others; being able to download, install and successfully link to an array of disparate software pieces on a given, often individualized, compute platform; having an effective pathway to communicate with the developers of the library for advice and to offer feedback and suggestions for improvement. These issues are not specific to the ESC but rather reflect generic challenges that confront all shared software development efforts. 

\subsection{\esl}

\subsubsection{Concept}

  This is where the ``Electronic Structure Library'' (\esl)\cite{esl-main} enters, 
the subject of the present paper.
  A key goal of the \esl is to alleviate and overcome the issues mentioned above,
  creating an effective collaboration platform for
shared software developments, where these make sense.
  Our vision is to sustain a community that develops, distributes and 
oversees 
electronic structure libraries for the benefit of all 
electronic structure codes.

  The \esl started in 2014 as a CECAM initiative, with the aim of stimulating
the segregation of well-defined tasks into shared libraries, pushing
the model of Fig.~\ref{fig:models}(b), and confronting the challenges
it entails.
  From the beginning, the work of the \esl has been done by programmers 
actively involved in successful electronic structure codes and the 
\esl initiative has been supported by the development teams of these codes, which include
ABINIT,\cite{gonze_abinit_2009} \siesta,\cite{siesta_soler_2002} 
\octopus,\cite{octopus_2020} Quantum ESPRESSO,\cite{giannozzi_qe_2017} \bigdft,\cite{Genovese2008_014109} \fhiaims,\cite{fhiaims_blum_2009} and \gpaw,\cite{mortensen_gpaw_2005} amongst others.

 The initial efforts focused on three aspects: 
($i$) identifying existing libraries suitable for inclusion in the \esl;
($ii$) extracting and re-coding as libraries a number of sub-packages 
from the community codes; and 
($iii$) incorporating these libraries into other participant codes.
 
Ongoing efforts within the \esl include improving the coordination between and 
interoperability of the various software modules, expanding their integration into 
large software development projects (e.g., some of the main electronic structure codes
in the community), and making it easier to seamlessly distribute a consistent 
bundle of library and software modules (see Sec.~\ref{sec:bundle}).

  A key enabler in all this process has been the will to 
overcome the monolithic mentality, both at a scientific level 
(one research group, one code) as well as at a business model level 
(free software vs. 
open-source vs. proprietary), allowing collaborations between 
communities and making new public-private partnerships possible.

  In addition to the obvious goal of avoiding re-inventing the 
wheel for every code, by re-coding well-known algorithms for 
well-established tasks, two other important advantages are 
foreseen. 
  The first relates to human resources.
  Electronic structure codes encompass  sophisticated physics
and sophisticated software engineering. The monolithic development
model demands highly educated personnel with expertise in both areas.
  An efficient segregation of tasks into libraries would allow an abstraction
of the low-level detail for physicists or chemists coding at a 
high-level, while software engineers could maintain and evolve the
low-level software without needing a high-level of expertise in the
science used.

  A related second advantage is that a widely-used \esl\ 
library or set of libraries, with well-defined APIs, 
would offer a good target for re-coding for software engineers 
working close to the cutting-edge of hardware developments and 
high performance computing (HPC) centres.
  This is, of course, a continuous process as it has to be 
done at each step of hardware evolution.
  Indeed, it has already happened with, e.g., Intel offering their
own implementation of linear-algebra libraries adapted to
their own compilers and processors. 
  The \esl should be able to offer many more targets for 
optimisation.
  It should also be remembered that there is currently a substantial level of resource dedicated to re-engineering
codes for new hardware, both at individual HPC centres
and funded by national (or trans-national, e.g., European Union)
research agencies.
  These efforts are usually directed towards particular codes.
  Dedicating these efforts to libraries would
be more efficient, serve the community more widely, 
and would also be easier to maintain as libraries are naturally composed of 
independent modules.
  Ideally, scientists should aspire to 
adapt their codes to new computers and to new computer paradigms as this is usually 
the only way to access the largest computational systems. Through the \esl, this could be achieved just 
by linking to the latest library implementation for 
a given computer architecture.
  
  These elements of the \esl vision rely, 
however, on the conversion into libraries of massively 
parallel heavy-duty code, which is an extremely ambitious
goal.
  There has been an emphasis on heavy-duty tasks in the \esl
efforts so far, although work has not focused exclusively on this.
  These efforts are described in Section~\ref{sec:implement}.
  The segregation of heavy-duty libraries involves many new 
challenges, which we now describe.

\subsubsection{Challenges}

  In addition to the challenges mentioned above referring
to shared software in general, the model proposed here faces a number of additional 
important challenges. Firstly, building a binary
code (compilation and linking), which depends on many libraries, and often their specific version, is substantially more difficult 
than for a self-contained (monolithic) program.
  Furthermore, the complexity of the heterogeneous 
environments typically encountered at HPC installations 
makes the build even harder and more diverse. 
  Ours is not the first community to face these problems,
and a significant part of the \esl effort is expended in the bundling
and building strategies for the \esl, as described in
Section~\ref{sec:bundle}.

  A second important challenge is the loss of the global
coherence in data structures and parallelisation that
monolithic programs can adopt (although it is not always 
possible or convenient).
  This implies the need for conversion routines to adapt data structures from one section of the code to another.
  Again, this challenge is not new, and it represents an intrinsic
element of this modular paradigm.
  Associated with these conversions, and in general, with the
whole strategy, is an expected loss of efficiency,
compared to that achievable within perfectly coded (and constantly
maintained) monolithic programs.
  However, the savings in (limited) human resources that modularity
brings are likely to outweigh a loss in (continuously expanding)
hardware cycles. An analogy can be made with the controversies in the
early seventies regarding the use of high-level languages (instead of
machine language) for the implementation of system software.\cite{C-Unix} It is now
clear that the apparently wasteful road led to significant
progress.

  Another challenge faced by the \esl to date stems from the fact
that the majority of the libraries currently in the \esl have been extracted 
from pre-existing electronic structure codes.
  This means that the API and internals of the library were chosen 
with its parent environment in mind. 
  Finally, the issue of licensing should be mentioned. 
  Different libraries are released under different licenses,
which may impose conditions on the licenses under which 
the using codes are distributed.
  This represents a challenge as well, although of a different 
kind.



\section{Common elements of electronic structure codes}
\label{sec:common}

Before describing the existing library implementations in the \esl, we give here a brief overview of the task that they are supposed to handle or, to put it more simply, what are the common elements of the electronic structure codes.
From the many available methods to approximate Dirac's equation in a
computationally tractable form, the majority fall into one of two broad classes:
density functional theory (DFT) and wave-function based methods.
  In this paper we concentrate on the former, although many of the tools
described here are also useful for other methods also based on effective
single-particle models, such as Hartree-Fock.

In the non-relativistic limit, the equations to be solved for ground-state DFT
are the Kohn-Sham equations:\cite{Martin2004}
\begin{equation}
  \hat{h}_\mathrm{KS}[n] \phi_i(\mvec{r}) = \epsilon_i \phi_i(\mvec{r})\,,
  \label{eq:ks}
\end{equation}
where $\phi_i$ and $\epsilon_i$ are the Kohn-Sham (KS) orbitals and eigenenergies,
respectively, and $\hat{h}_\mathrm{KS}$ is the Kohn-Sham hamiltonian. The
hamiltonian is usually decomposed in the following way:
\begin{equation}
  \hat{h}_\mathrm{KS}[n] = \hat{t}_s + v_\mathrm{ext} + v_\mathrm{H}[n] + v_\mathrm{xc}[n]\,,
  \label{eq:hks}
\end{equation}
where $v_\mathrm{ext}$ is the external potential (typically the potential
generated by the nuclei), $v_\mathrm{H}$ is the Hartree potential, and
$v_\mathrm{xc}$ is the exchange and correlation potential. $\hat{t}_s$ is the single-particle kinetic energy operator. In non-relativistic form, 
\begin{equation}
    \hat{t}_s = -\frac{1}{2}\bnabla^2 .
\end{equation}
However, practically every electronic structure code employs at least a scalar-relativistic variant of $\hat{t}_s$ (the applicability of the non-relativistic expression is limited to the lightest chemical elements only). In codes employing pseudopotential-type techniques (see below), relativity is usually incorporated implicitly through the form of the projectors. In all-electron codes, explicit scalar-relativistic forms of $\hat{t}_s$ are used.
The Kohn-Sham
equations are a set of one-particle equations that need to be solved
self-consistently as several terms in Eq.~\eqref{eq:hks} are functionals of the
electronic density:
\begin{equation}
  n(\mvec{r}) = \sum_i f_i |\phi_i(\mvec{r})|^2\,.
  \label{eq:rho}
\end{equation}
$f_i$ are occupation numbers, ensuring that the orbitals are only occupied as far as there are electrons (i.e., $\sum_i f_i = N_\mathrm{el}$, where $N_\mathrm{el}$ is the number of electrons in the system). Any code that aims to solve the Kohn-Sham equations must therefore perform the
following tasks: 1. given a set of atomic coordinates, evaluate
$v_\mathrm{ext}$; 2. evaluate $v_\mathrm{H}[n]$; 3. evaluate $v_\mathrm{xc}[n]$;
4. solve the eigenvalue problem of Eq.~\eqref{eq:ks}; 5. find the density that
solves the self-consistency problem. Each of these steps thus represents an
opportunity for electronic structure packages to share and reuse code:

\begin{enumerate}
  \item To reduce the computational cost, many DFT codes use the pseudopotential
    approximation.~\cite{Hamann2013} The pseudopotentials are normally generated by
    specialized codes that output them using a particular one of the existing file
    formats. Therefore, codes that want to use a specific pseudopotential are required to
    know the corresponding file format to parse the corresponding information.
  \item The Hartree potential $v_\mathrm{H}[n](\mvec{r})$ is defined as
  \begin{equation}
  \label{eq:hartree1}
    v_\mathrm{H}[n](\mvec{r}) = \mint{\mvec{r}'} \frac{n(\mvec{r}')}{|\mvec{r} - \mvec{r}'|}\,.
  \end{equation}
  Direct evaluation of this integral is not usually numerically efficient
  and it is common practice to instead solve the corresponding Poisson equation.
  \item Many hundreds of different approximations to the exchange-correlation
    functional have been proposed, some of which require the evaluation of
    long, complex mathematical expressions. Implementing such approximations is
    thus a tedious, error prone task.
  \item Many different methods exist in the literature for solving eigenvalue
    problems such as Eq.~\eqref{eq:ks}. Upon discretization of the
    orbitals $\phi$, one can write the problem in the language of matrices and
    vectors. Then solving Eq.~\eqref{eq:ks} reduces to the standard linear algebra
    problem of diagonalizing a matrix, in this case the hamiltonian matrix. For
    cases where the size of this matrix is too large for direct diagonalisation, 
    either due to the memory or computational time required, iterative eigensolvers 
    can be used which only require the result of the hamiltonian operating on an
    orbital, or alternative formulations of the problem can be solved, such as
    the ones based on Green's functions or Fermi-operator expansions.

  \item Finding the density that solves the self-consistency problem is usually
    done iteratively: starting from a guess for the density, one solves
    Eq.~\eqref{eq:ks}, thus obtaining a new set of orbitals $\phi$ which, in
    turn, are used to obtain a new density. The process is then repeated using the
    new density until the changes in the density are smaller than some 
    defined threshold. Since the total computational cost strongly depends on
    how fast the iterative procedure converges, many methods are available to
    accelerate this process.
\end{enumerate}

In the case of wave-function based methods, many of these require 
a solution to the Hartree-Fock equations as a starting point. These equations
share many similarities with the Kohn-Sham equations: both are sets of
one-particle equations that need to be solved self-consistently. This further
increases the opportunities for code sharing and reuse among electronic
structure packages.

To numerically solve either the Kohn-Sham or the Hartree-Fock equations, the
relevant quantities are typically discretized in some way, either using
basis-sets or grids. Each type of discretization requires specialized functions
that can, in principle, be shared among codes that use the same basis-set or type
of grid. For example, atom-centred basis sets require the efficient evaluation
of one- and two-particle integrals.

Along with the common elements of electronic structure packages 
which are directly related to the equations solved, other types of 
operations are also performed by most ES codes.
A prime example are I/O operations, which range from parsing an input file 
to writing physical quantities of interest to disk for
visualization or further processing.

\section{Existing library implementations in the \esl}
\label{sec:implement}

In the following, we briefly present the libraries and packages that are
currently part of the \esl, giving a brief description of their scope,
history, and use cases. They are tabulated in Table~\ref{tab:libs}.

\begin{table*}[tb]
\centering
\caption{\label{tab:libs}
Libraries included in the \esl and \esl bundle. The first twelve are dedicated to
electronic structure (ES) functionality; the last ten are tools of more general (GEN) applicability
beyond electronic structure theory. They are described in Section~\ref{sec:implement}, except
for Futile, which is described in Section~\ref{sec:bigdft}.
  The licence acronyms expand as follows:
GPL: GNU General Public Licence, in versions 2.0\cite{gpl2} and 3.0;\cite{gpl3} 
LGPL: GNU Lesser GPL, version 3.0;\cite{lgpl3}
MPL: Mozilla Public Licence, version 2.0;\cite{mpl2}
MIT: Massachusetts Institute of Technology Licence;\cite{MITlicence}
CeCILL-C: The CeCILL-C Free Software License Agreement;\cite{cecillc}
BSD: Berkeley Software Distribution licence, in
either the 2-clause\cite{bsd2} or the 3-clause\cite{bsd3} versions.}
\begin{ruledtabular}
\begin{tabular}{lll}
Library & \multicolumn{1}{c}{Functionality} & Licence  \\ 
\hline
\psolver & ES: Poisson solver for 0, 1, 2 and 3 dimensions, varying dielectrics 
               and Poisson-Boltzmann & GPL-2.0 \\
\libxc  & ES: Pointwise evaluation of exchange \& correlation for LDAs and GGAs & MPL-2.0   \\
\libvdwxc & ES: Evaluation of Van der Waals non-local exchange \& correlation  & GPL-3.0 \\
\libgridxc & ES: Evaluation of exchange \& correlation in regular grids incl. 
                  non-local Van der Waals DFs & BSD 3-clause \\
\pspio & ES: Input/output of pseudopotentials in most popular formats & LGPL-3.0 \\
\libpsml & ES: Standardized pseudopotential markup language specification 
                and associated library & BSD 3-clause \\
\escdf & ES: Electronic-structure data format specification and associated library & LGPL-3.0\\
\elsi & ES: Unified interface calling a variety of Hamiltonian solver 
        libraries& BSD 3-clause \\
\pexsi & ES: Pole expansion and selective inversion solver library & BSD 3-clause \\
\libomm & ES: Iterative minimization non-orthogonal solver & BSD 2-clause\\
PIKSS & ES: Parallel iterative Kohn-Sham solvers & GPL-3.0 \\
\wannier  &  ES: Postprocessing to obtain maximally-localized Wannier functions and
                       derived quantities & GPL-2.0\\
\elpa & GEN: High-performance dense eigenvalue solver library & LGPL-3.0 \\
\ntpoly & GEN: Sparse linear-scaling solver library & MIT \\
\sips & GEN: Shift-and-invert parallel slicing solver & BSD 2-clause \\
SuperLU\_DIST & GEN: Sparse linear system solver & BSD 3-clause \\
Scotch & GEN: Graph partitioning library & CeCILL-C \\
\matrixswitch & GEN: Matrix-format-independent 
                         abstraction layer of linear algebra operations & BSD 2-clause\\
\flook & GEN: Connection between Fortran and Lua for embedded scripting code control& MPL-2.0\\
\libfdf & GEN: Flexible data format for input of control parameters & BSD 3-clause \\
xmlf90 & GEN: Fortran library to parse and write well-formed XML files & BSD 2-clause \\
Futile & GEN: Low-level toolbox (handles YAML-code mapping, dynamic memory, timing, 
         error, etc.) & GPL-3.0\\
\end{tabular} 
\end{ruledtabular}
\end{table*}

\subsection{\psolver}
 \label{section:Hartree}
Electrostatic potentials play a fundamental role in nearly any field of physics and chemistry.
It is, therefore, essential to have efficient algorithms to find the electrostatic potential $V$ arising from a charge
distribution $\rho$ (associated to the particle density $n$ in
Eqs.~\eqref{eq:rho} and \eqref{eq:hartree1}) in a dielectric medium
described by the dielectric constant $\epsilon( \textbf{r})$, or, 
in other words, to solve the generalized Poisson's equation
\begin{equation}
\label{GPe}
\nabla \cdot \epsilon( \textbf{r}) \nabla \phi(\textbf{r}) = -4 \pi \rho(\textbf{r}) .
\end{equation}

The large variety of situations in which this equation is encountered led us to address this
problem for different choices of the boundary conditions (BC).
The long-range behavior of the inverse Laplacian operator makes this problem strongly dependent on
the BC of the system. Therefore, any method aiming at providing a solution to Eq.~\eqref{GPe} has to
deal with the BC, which, for instance,  
could be either periodic or free (otherwise referred to as ``isolated'' or ``open'') along each of the three directions $x,y,z$. In the case of fully periodic BC, the most natural (and efficient) approach to the problem is the reciprocal space treatment. It amounts to expanding both the density and the potential as superpositions of plane waves (Fourier series), thereby Eq.~\eqref{GPe} becoming -- for a homogeneous dielectric -- algebraic in the Fourier components of $\rho$ and $V$.
This equation is readily solved and the result is finally transformed back into real space. Forward and backward transformations are carried out via Fast Fourier Transforms (FFT), hence the overall computational scaling of the method with respect to the number $N$ of grid points is a rather appealing $\mathcal{O}(N\log N)$.

The situation is less straightforward for the same problem but different BC, e.g., free (isolated) BC.
In this case the solution of Poisson's equation in vacuum can formally be obtained from a three-dimensional integral:
\begin{equation} \label{greenequation}
V(\mathbf r) =\int d \mathbf r' G(| \mathbf r - \mathbf r' |) \rho(\mathbf r')\;,
\end{equation}
where $G(r)=1/r$ is the Green function of the Laplacian operator in the unconstrained $\mathbb{R}^3$ space.
The long range nature of the kernel operator $G$ does not allow us to approximate free BC with a very large periodic volume.  Consequently, the description of non-periodic systems using a periodic formalism always introduces long-range interactions between supercells that compromise the results.

Due to the simplicity of plane wave methods, various attempts have been made to generalize the reciprocal space approach to free BC.~\cite{Hockney.1970,Fusti-Molnar-Pulay.2002,Martyna-Tuckerman.1999} All of them use a FFT at some point, and thus have a $\mathcal O(N \log N)$ scaling.
These methods use \textit{ad hoc} screening functions to subtract the spurious interactions between super-cells. They have some restrictions and cannot be used blindly. For example, the method of F\"usti-Molnar and Pulay\cite{Fusti-Molnar-Pulay.2002} is only efficient for spherical geometries and the method of Martyna and Tuckerman\cite{Martyna-Tuckerman.1999} requires artificially large simulation boxes that are computationally expensive.
Nonetheless, the usefulness of reciprocal space methods has been demonstrated for a variety of applications, and plane-wave based approaches are widely used in the chemical physics community.

  Two-dimensional periodic systems, such as surfaces, are another prominent choice of BC.
  The many surface-specific experimental techniques developed in recent years produce important results 
that can greatly benefit from theoretical interpretation and analysis.
  The development of efficient computational techniques for systems with such 
boundary conditions thus became very important.
A number of explicit Poisson solvers have been developed in this framework~\cite{TuckermanSurfaces, HockneySurfaces, MortensenSurfaces} based on a reciprocal space treatment. Essentially, these Poisson solvers are constructed by implementing a suitable generalization for surface BC of the same methods that were developed for isolated systems. As for the free BC case, screening functions are applied to subtract the artificial interaction between the supercells in the non-periodic direction.
Therefore, they exhibit the same kind of intrinsic limitations, e.g., good accuracy is only achieved inside the bulk of the 
computational region, with the consequent need for artificially large simulation boxes, 
which may increase the computational overhead.

Following these considerations, 
a series of efficient and accurate Poisson solvers have been developed that compatible with all possible combinations of mixed isolated/periodic boundary conditions. The solvers also support screened and unscreened Coulomb operators in vacuum~\cite{PSfree,PSsurface,PSwires} and distributed, non-uniform dielectrics including the Poisson-Boltzmann equation.~\cite{PSgeneralized,PSSoftSphere}
In contrast to Poisson solvers based solely on a reciprocal space treatment,
the fundamental operations of this Poisson solver are based on a mixed reciprocal-real space
representation of the charge density. This allows different boundary conditions in different directions to be naturally satisfied. Screening functions or other approximations are thus not needed.

The basic advantage of this approach is that the real-space values of the potential 
$V(\mathbf{r})$
are obtained to very high accuracy on the uniform mesh of the simulation domain, via a direct solution of
Poisson's equation by convolving the density with the appropriate Green's function of the Laplacian. As already mentioned, the Green's function can be discretized for the most common types of boundary
conditions encountered in electronic structure calculations, namely free, wire, slab and periodic.
This approach can therefore be straightforwardly used in all DFT codes that are able to express
the densities 
$\rho(\mathbf{r})$
on uniform real-space grids.
This is very common because the XC correlation potential is usually calculated on such a grid, at least in pseudopotential-based codes.
This approach has also proved to be, in its parallel CPU version, the fastest in most cases~\cite{garcia2014survey} and is therefore integrated in various DFT
codes such as {\sc abinit},~\cite{gonze_abinit_2009} {\sc CP2K},~\cite{hutter2014cp2k}
\octopus,~\cite{andrade2012time,octopus_2020} and {\sc Conquest}.~\cite{CONQUEST_2007}

To conclude, the Poisson solver algorithm has already been ported on Graphic Processing Units (GPU)~\cite{PSGPU} and is readily available in the \esl package.
It enables affordable calculation of exact exchange operators in large systems.~\cite{ExactExchange}

\subsection{\libxc}\label{sec:libxc}
  The exchange-correlation functional is at the heart of density-functional 
theory,\cite{Martin2004} and it is ultimately responsible for the 
accuracy of any such electronic structure calculation. 
  It is, therefore, perhaps not surprising that hundreds of different approximations to this term have been proposed over the last decades. Most of these can be classified into five families, usually often identified as different rungs of Jacob's ladder,~\cite{Perdew2001_1} leading from the Hartree world to the Heaven of chemical accuracy. The rungs correspond to the local-density approximation, the generalized-gradient approximation, the meta-generalized-gradient approximation, functionals that depend on the occupied Kohn-Sham orbitals, and finally, functionals that also depend on the virtual orbitals. \libxc~\cite{Marques2012_2272,Lehtola2018_1} is a library that contains the mathematical expressions for functionals belonging to the first three families, together with the semi-local parts for the functionals of the last two rungs.
 
 \libxc has, by now, a long history, with its roots at the beginning of this century and version 1.0.0 appearing in 2010 (the current stable version is 4.3.4). The number of functionals included has increased steadily over the years with more than 500 functionals, arising from more than half a century of theoretical developments, implemented to date. Recently, the library was completely restructured to allow the definition of the functionals to be written in Maple 2016 (Ref.~\onlinecite{maple}), which simplifies the insertion of new functionals (Maple's symbolic language is considerably simpler than C, and well adapted for mathematical manipulations).
  Moreover, all derivatives are evaluated symbolically by Maple. 
  This significantly reduces the possibility of errors in the implementation and opens the way for the evaluation of higher derivatives of the functionals. Currently, \libxc supports up to fourth-derivatives, required, for example, for the calculation of Hessians of potential energy surfaces for excited-states. 
 
 There are a number of advantages of \libxc for the users of electronic structure codes. First, they have instant access to nearly all the exchange-correlation functionals ever developed. Furthermore, most functionals are implemented in \libxc shortly after their publication, giving access to the latest theoretical developments in density-functional theory often only requiring a simple recompilation of the library. Finally, it makes the comparison of different codes and methods much simpler. \libxc is by now used by more than 30 electronic structure codes, developed both by the Physics communities (such as Abinit,~\cite{Gonze2016_106} \bigdft,~\cite{Genovese2008_014109} \fhiaims,~\cite{Blum2009_2175} WIEN2k,~\cite{Blaha2018} etc.), the Quantum Chemistry community (such as Psi4,~\cite{Parrish2017_3185} Orca,~\cite{Neese2012_73} PySCF,~\cite{Sun2018_e1340} or Turbomole,~\cite{Ahlrichs1989_165} etc.), commercially developed codes (QuantumATK~\cite{Smidstrup_2019}), as well as other libraries, e.g., \libgridxc [see Sec.~\ref{sec:gridxc}]. \libxc guarantees reliable, bug free implementations of the functionals, which are often cross-checked with reference code from the original authors of the functionals. Finally, \libxc provides a simple means to perform benchmark calculations in a variety of physical systems and using diverse numerical methods (see, e.g. Ref.~\onlinecite{Borlido2019_5069}).

\subsection{\libvdwxc}
\newcommand{\askhl}[1]{{\textcolor{red}{[askhl (asklarsen@gmail.com): #1]}}}
\label{sec:vdwxc}

 \libvdwxc~\cite{libvdwxc-paper} is a software library which evaluates the
the non-local correlation term for density functionals in the vdW-DF
family~\cite{BerCooLee15,HylBerSch14} such as
vdW-DF,~\cite{langreth05p599} vdW-DF2,~\cite{langrethjpcm2009}
and vdW-DF-cx.~\cite{BerHyl14}
It also implements the recent spin-generalization
of these functionals.~\cite{Thonhauser_2015:spin_signature}
It is written in C and released under the GNU GPL licence.

\libxc evaluates functionals point-wise and hence supports
only local and semi-local functionals.
  The purpose of \libvdwxc is to complement
\libxc by providing just the missing non-local term.
\libgridxc contains an alternative implementation  
--- see Section \ref{sec:gridxc} below.

The vdW-DF functionals are the sum of three terms: The correlation energy from
LDA; the exchange energy from a GGA functional, which is often chosen
differently for different vdW functionals; and finally
the non-local vdW correlation energy
which is characteristic of the vdW-DF family.
This latter term is an integral over a kernel
function $\phi(\mathbf r, \mathbf r')$:
\begin{align}
  E_{\mathrm{c}}^{\mathrm{nl}}[n] =
  \frac12
  \iint
  n(\mathbf r)
  \phi(\mathbf r, \mathbf r')
  n(\mathbf r')
  \, \mathrm d \mathbf r \, \mathrm d \mathbf r'.
\end{align}
Direct integration of this expression scales as $\mathcal O(N^2)$ and is
very expensive, so most codes use the
spline interpolation method due to Román-Pérez and Soler.~\cite{vdw_RomanPerez_Soler_2009}
This reduces the
integral to a convolution in Fourier space
whose computational cost is only $\mathcal O(N \log N)$.

The algorithm uses a number (conventionally 20) of helper
functions,~\cite{vdw_RomanPerez_Soler_2009} $\theta_n(\mathbf r)$, and their
Fourier transforms.  This still requires more memory and computation time
than a standard GGA functional.  \libvdwxc focuses on parallel scalability in order
that this computation will not become a bottleneck.  It works in parallel using
MPI with the Fourier transform library FFTW.~\cite{FFTW05,JohnsonFr08:burrus}
For parallel computations, the grids use the 1D block distribution of FFTW.
\libvdwxc additionally supports the PFFT library,~\cite{pfft-pippig} an extension
to FFTW which improves scalability for massively parallel architectures.

\libvdwxc takes the density and its gradient on a uniform 3D grid as
input. The grid directions need not be orthogonal.
It calculates the total energy and its derivatives at each
point, following \libxc conventions
for ease of integration with DFT codes.

\subsection{\libgridxc}
\label{sec:gridxc}

The \libgridxc library\cite{weblibgridxc} started
life as SiestaXC, a collection of modules within \siesta to compute the
exchange-correlation energy and potential in DFT calculations for
atomic and periodic systems. The ``grid'' part of the name refers to
the discretization for charge density and potential used in those
calculations. The original code included a set of low-level routines to
compute the exchange-correlation energy density and potential,
$\epsilon_{xc}({\bf r})$ and $V_{xc}({\bf r})$, respectively, 
at a point for (semilocal) LDA and GGA functionals (i.e.,
a subset of the functionality now offered by \libxc), and two
high-level routines to handle the computations in the whole domain
(with radial or 3D-periodic grids), including  computations
of any gradients, integrations, etc, needed. The most relevant feature of SiestaXC
was its pioneering
implementation of efficient and practical algorithms for 
van der Waals functionals,~\cite{vdw_RomanPerez_Soler_2009} in
particular for the evaluation of the non-local correlation term. These
algorithms have found their way into numerous other implementations, as
exemplified in Sec.~\ref{sec:vdwxc} on \libvdwxc. Another strength of the
code is its support for non-collinear spin densities, as needed in
particular for calculations with spin-orbit-coupling.
Like \libvdwxc, it inputs the density on a uniform grid, not 
necessarily orthogonal, and outputs the XC energy and potential 
on the same grid. But in contrast with \libvdwxc, the density 
gradient is evaluated internally.

The current \libgridxc retains most of the SiestaXC functionality, and
enhances it by offering an interface to \libxc that supports a much
wider selection of XC functionals. The code, written in Fortran, has been streamlined and
re-packaged into a proper stand-alone library, with an automatic
build-system. It is used by modern versions of \siesta, it is being 
adopted in \abinit, and it is also being
considered for \bigdft and other codes.

\subsection{\pspio}
\label{sec:pspio}

For a long time, the development of pseudopotentials has generally been coupled to a parent DFT code. This has resulted in a proliferation of file formats and incompatibilities, preventing or severely limiting collaboration involving different codes. Even worse, some versions of a pseudopotential format are not compatible with some versions of the DFT code they originated from. To address this issue, many discussions took place from 2002 on to define a common file format for pseudopotentials. While this led to the successful creation of the PAW-XML format for projector augmented-wave (PAW) datasets,~\cite{paw_xml_specs} 
no agreement was reached at the time for norm-conserving pseudopotentials
(see, however, Section~\ref{sec:psml} below).

\pspio takes exactly the reverse perspective: since many file formats exist and will continue to exist for the foreseeable future, let us design and implement a library that is able to read and write all of them, including the different versions of each format. Any pseudopotential generator or DFT code using \pspio will thereby be free of  file-format problems. However, \pspio is not intended to act as a ``universal translator'', which would basically require implementation of a pseudopotential generator within the library. Indeed, different file formats store different quantities, some of which have to be reconstructed to convert one format to another. As a consequence, direct format conversion is only possible in a very limited number of cases.

\pspio currently supports the \texttt{FHI98PP}, \texttt{ABINIT6} and \texttt{UPF-1} file formats. Support for \siesta \texttt{PSF} and \texttt{ONCV} formats is currently being tested.
It can be found in Ref.~\onlinecite{pspio}.

\subsection{\libpsml}
\label{sec:psml}

Several well-known programs generate pseudopotentials in a variety of
formats, tailored to the needs of specific electronic-structure codes. While
some generators are now able to output data in different bespoke
formats, and some simulation codes are now able to read different
pseudopotential formats (with the help from \pspio in Section~\ref{sec:pspio}, 
for example), the common historical pattern in the design
of those formats has been that a generator produced data for a single
particular simulation code, most likely to be the one maintained by the same
group. The consequence was often that a number of implicit assumptions, shared by
generator and user, have entered into the formats and fossilized there.

This leads to practical problems, not only of programming, but also of
interoperability and reproducibility, which depend on spelling out a large number of details which are not always well known or documented for all codes or existing formats.

PSML (for PSeudopotential Markup
Language)~\cite{psml_Garcia_2018,psml_info} is a file format for
norm-conserving pseudopotential data which is designed to encapsulate, to the greatest extent possible, the abstract concepts in the domain's ontology,
and to provide appropriate metadata and provenance information. PSML
files can be produced by the {\sc ONCVPSP}~\cite{Hamann2013} and {\sc
  ATOM}~\cite{Froyen} pseudopotential generator programs, and are a
download-format option in the Pseudo-Dojo database of curated
pseudopotentials.~\cite{dojo_vanSetten_2018,pseudo-dojo-site}

The software library \libpsml~\cite{psml_Garcia_2018,psml_info} can be
used by electronic structure codes to transparently extract the
information in a PSML file and adapt it to their own data structures,
or to create converters for other formats. It is currently used by
\siesta and \abinit, making full pseudopotential
interoperability possible and thus facilitating comparisons of calculation results.

  A feature of the PSML format and library is worth noting: the
exchange-correlation flavor used in the generation of the
pseudopotential is encoded in the PSML file as a set of \libxc
``ids''. 
  It exemplifies the importance of software standards in scientific
computing and their implementation in widely available libraries.
  Given the comprehensive support for functionals in \libxc,
this is very close to a ``universal'' specification.
  The combination of \libpsml and \libxc (with maybe \libgridxc 
as an intermediate layer) is thus a basic ingredient for interoperability.

\subsection{Electronic structure common data format (\escdf)}

The electronic structure common data format (\escdf)~\cite{escdf_web} and the accompanying library
\libescdf~\cite{libescdf_web} are currently being developed with the aim of simplifying a number of 
I/O related issues: (1) many codes deal with the same information, foremost structural data about 
the system of interest, which could easily be interchangeable between codes, and for which a common
format would be useful; (2) having a common standard available would simplify workflow systems,
chaining e.g. ab initio calculations with post-processing spectroscopy calculations and data 
visualization; (3) parallel I/O of large data sets for general output or for code-specific restart
files is becoming increasingly important and having a common tool to facilitate this at a low level
would help many code developers.
Over the last decades, there have been several attempts to introduce such common standards in the ESC, with
varying degrees of success. The main challenge is that much of the data is not actually interchangeable between codes which are based
on different computational methods. For instance, it is, in general, not meaningful to use wave
functions generated in a plane wave pseudo-potential method in an all-electron LAPW code. \escdf 
acknowledges that fact and does not try to impose a rigid standard but, rather, to provide lower 
level tools which define a common vocabulary for writing data and to provide the necessary
meta-data to clearly describe how the data in a given file is represented.
  This ambition for flexibility is further illustrated in the specification of 
structure, which allows for periodicity in any dimension (0 to 3), as used by
the \psolver (Section~\ref{section:Hartree}), and even beyond, as for non-periodic embedding
in infinite or semi-infinite structures, as used by multiple-scattering
methods.\cite{Sebilleau2011,Ebert2011,Ebert2016} 

The ideas behind the \escdf are, to a large extent, based on the ETSF-IO library and 
associated specifications,~\cite{etsf_web,etsf_specs} which it tries to extend and 
modernize by moving from netCDF-4 to HDF-5~\cite{hdf5_web} as the underlying technology. 
They also build on the wavefunction format of the BerkeleyGW code,\cite{BerkeleyGW} 
which was defined as both a specification and a library of reading and writing 
routines, used by \octopus and other DFT codes in preparing inputs for $GW$ and 
Bethe-Salpeter calculations.
The development has been
driven by a collaboration of ETSF developers and new developers, in particular from the 
EUSpec~\cite{euspec_web} network, which had the specific goal of providing tools for chaining
calculations and post-processing tools. The \escdf specifications also have been aligned as much
as possible with existing specifications from the NOMAD project~\cite{nomad_web} and have also informed NOMAD
about their new extensions.
The \escdf specifications are developed and maintained by a 
dedicated curating team. 

One of the core features of the implementation of \libescdf is the separation of the format
specification and the library code. The specifications are defined in a JSON file, which can easily
be extended without the need to change the code in the library. Specific code for the library is
then generated automatically from the JSON file and the format documentation is also auto-generated
from this central specifications file. This strategy will make the library more maintainable and
effectively decouples the science from the underlying software design.

Currently, both the specifications and the library are still under development, with several sections
of the former already complete. As soon as it is possible, \libescdf will be interfaced with the \esldemo project
and included in the \eslbundle (see Sec.~\ref{sec:bundle}).

\subsection{\elsi and supported solver libraries:
\elpa, \pexsi, \ntpoly, \sips, SuperLU-DIST, Scotch}
\label{sec:elsi}

\begin{figure}[!htbp]
\centering
\includegraphics[width=0.45\textwidth]{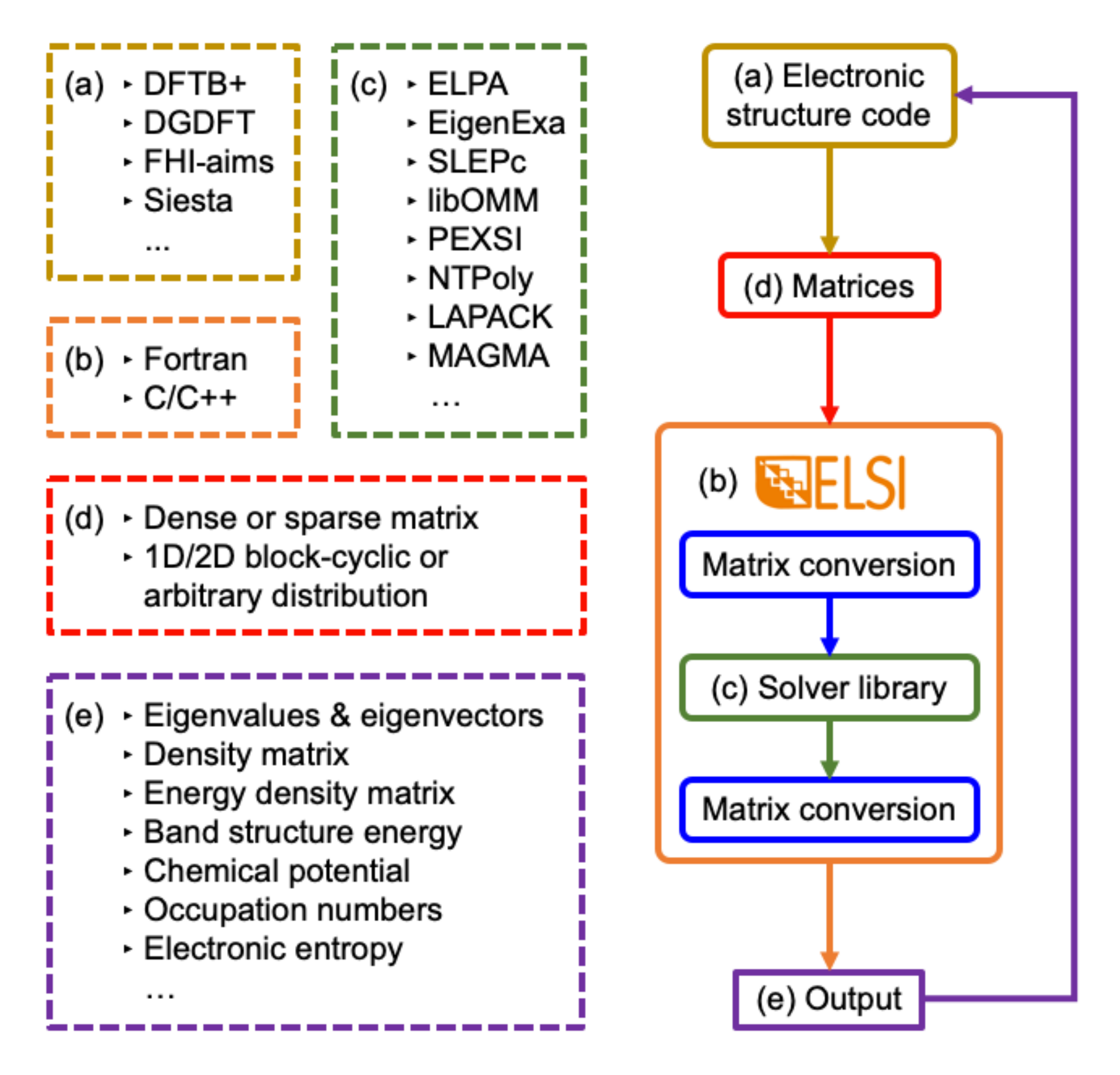}
\caption{Interaction of the \elsi interface with electronic structure codes. An electronic structure code has access to various eigensolvers and density matrix solvers via the \elsi API. Whenever necessary, \elsi handles the conversion between different units, conventions, matrix formats, and programming languages. 
  (a) lists the electronic structure codes that currently use \elsi. 
  (b), (c), (d), and (e) list the programming language, solvers, matrix 
formats, and output quantities, respectively, supported by \elsi.}
\label{fig:elsi}
\end{figure}

This group of libraries solves or circumvents the Kohn-Sham or generalized Kohn-Sham eigen problem, i.e. the central problem of electronic structure calculations. They
can be used in conjunction with the open-source \elsi library (ELectronic Structure Infrastructure, 
\url{https://elsi-interchange.org}), but the associated solvers can be and are also used in a standalone fashion with different electronic structure packages. ELSI provides a unified software interface that connects electronic structure codes to various high-performance solver libraries to solve or circumvent eigenproblems encountered in electronic structure theory.~\cite{elsi_yu_2018} In addition to providing interfaces, matrix conversion, etc., ELSI also abstracts common tasks in handling eigenvalue problems in an electronic structure code.
The tasks handled by \elsi and related solvers often amount to the most compute-intensive ones in electronic structure codes. These ESL components therefore already offer support for several past and present pre-exascale hardware developments, notably Intel's many-core architectures as well as NVidia's GPUs.
Solvers currently supported in \elsi include conventional dense eigensolvers (\elpa,~\cite{elpa_marek_2014,PKus19_article} EigenExa,~\cite{eigenexa_imamura_2011} LAPACK,~\cite{lapack_anderson_1999} and MAGMA~\cite{magma_dongarra_2014}), the orbital minimization method (\libomm~\cite{libomm_corsetti_2014}), sparse iterative eigensolvers (SLEPc~\cite{slepc_hernandez_2005} and \sips), the pole expansion and selected inversion method (\pexsi~\cite{pexsi_lin_2013}), and linear scaling density matrix purification methods (\ntpoly~\cite{ntpoly_dawson_2018}). As sketched in Fig.~\ref{fig:elsi}, an electronic structure code interfacing to \elsi automatically has access to all the eigensolvers and density matrix solvers supported in \elsi. In addition, the \elsi interface is able to convert arbitrarily distributed dense and sparse matrices to the specification expected by the solvers, taking this burden away from the electronic structure code. A comprehensive review of the capabilities of the latest version of \elsi, including parallel solution of problems found in spin-polarized systems (two spin channels) and periodic systems (multiple \textbf{\textit{k}}-points), scalable matrix I/O, density matrix extrapolation, iterative eigensolvers in a reverse communication interface (RCI) framework, is presented in a separate publication.~\cite{elsi_yu_2019}

The development of \elsi, including its API design, internal data structure, build system, testing, and integration with electronic structure codes, was driven from its inception by contributions and feedback from the community. In workshops organized by the \esl, \elsi has been a primary focus from the outset. Moreover, developers and users of several electronic structure codes participate in open \elsi monthly video meetings to exchange ideas, ensuring a direct information flow in order to develop \elsi as a software package that fits the needs of as many electronic structure projects as possible. To date, the \elsi interface has been adopted by the DFTB+,~\cite{dftb-plus-new} DGDFT,~\cite{dgdft_hu_2015} \fhiaims,~\cite{fhiaims_blum_2009} and \siesta~\cite{siesta_soler_2002} codes. To aid in the selection of the solver that is best suited for a particular application problem, \elsi provides a series of benchmarks to assess the performance of the solvers for different problem types and on different computer architectures.~\cite{elsi_yu_2018,elsi_yu_2019} This benchmark effort has been greatly accelerated by a separate FortJSON library, shipped with \elsi, which enables the output of runtime parameters, matrix dimensions, timing statistics etc. into a standard JSON file. Thanks to the popularity and portability of JSON, \elsi log files written by FortJSON can be easily processed and analyzed by existing tools. Comparing different solvers in different codes on an equal footing is thus significantly simplified by the \elsi infrastructure.

  \elsi ships with its own tested versions of several individual solver 
libraries (which are also included in the \esl) but, additionally, linking against already compiled upstream 
versions from each solver library is supported as much as possible. 
  The installation of the different components is handled by a single 
CMake-based build system that either compiles redistributed source 
code of the solvers or links \elsi against user-supplied solver libraries.

\subsection{\libomm non-orthogonal eigensolver}
\label{sec:omm}

Now integrated into the larger bundle of eigensolvers provided by \elsi, \libomm was developed during the first \esl workshop as a standalone library, and can still be used as such. It is written in Fortran with C bindings, and can be compiled either for serial or MPI parallel operation.

The orbital minimization method (OMM) is an iterative solver method based on finding the set of Wannier functions describing the occupied subspace by the minimization of a specially-defined energy functional. The peculiarity of the OMM is that only an unconstrained minimization is required, thus avoiding a potentially expensive orthogonalization step; the properties of the functional drive the Wannier functions towards orthonormality as it is minimized. The OMM has an interesting history (discussed briefly in Ref.~\onlinecite{libomm_corsetti_2014}) stemming from research on linear-scaling DFT methods. The \libomm library, however, is based on a later re-implementation of the method used in \siesta as an efficient cubic-scaling solver for a basis of finite-range numerical atomic orbitals.~\cite{libomm_corsetti_2014}  As such, the library provides a tensorial correction for non-orthogonal basis sets, either via a Cholesky factorization of the overlap matrix or a preconditioner suitable for localized orbitals.

The library is built for maximum efficiency in data reuse; the API is designed for repeated calls within an outer self-consistency loop in the host code. Data is reused between calls in two ways: (a) some matrices, such as the coefficients matrix of the Wannier functions, are repeatedly passed in at each call, updated during the call and passed out again; (b) other data is allocated and stored internally by the library, and, therefore, a final call to free all memory must be performed.

The library is also written to be agnostic with respect to both the data storage scheme of the matrices and the implementation of the matrix operations. This is achieved by making use of an underlying library for matrix operations, \matrixswitch, described in Section~\ref{matrixswitch}.


\subsection{PIKSS: Parallel iterative Kohn-Sham solvers}

 In addition to ELSI and its supported solvers (Secs.~\ref{sec:elsi}
and \ref{sec:omm}), the parallel iterative 
Kohn-Sham (KS) solvers library PIKSS is a bundle of several iterative diagonalization
eigensolvers 
that have been extracted from the Quantum ESPRESSO (QE) suite, and recast in an independent, 
code-agnostic fashion.
  It includes the popular Davidson diagonalization and a band-by-band conjugate gradient 
minimization methods, but also implements the more recently developed Projected 
Preconditioned Conjugate Gradient (PPCG),\cite{ppcg2015} and Parallel Orbital (ParO) update 
solvers\cite{paro2017} that allow new parallelization paradigms. 

  As ideal within the \esl concept, the library is designed such that the 
interaction with the main electronic structure code is via routine library 
calls with a well-specified API.
  The operations performed by the library depend on the chosen diagonalization method, 
but generally include application of the hamiltonian to a set of candidate wavefunctions, 
computation of the overlap matrix, approximation of inverse matrices, etc.  

  Unlike the original, strictly plane-wave implementation in QE, the KS-Solvers library 
allows for any internal representation of the wavefunction and Hamiltonian.
  To further exemplify how to expand the usability of the solvers, a Reverse 
Communication Interface version for one of the solvers (Davidson diagonalization) is also provided. 
  The library is currently hosted in Ref.~\onlinecite{PIKSS}. 

  Initially integrated within KS-Solvers but currently being developed as a stand-alone library, 
the miniPWPP module serves as a demonstrator for the KS-Solver library. 
  miniPWPP, a barebone DFT implementation based on planewave-empirical pseudopotential framework, 
showcases the usage of the several methods within KS-Solvers library.
  It allows performance comparison of them on different hardware platforms 
(e.g. CPU, GPU), and with different parallelization paradigms (e.g. over bands, 
task groups etc.). 
  Both the KS-Solvers and miniPWPP, along with other libraries (i.e. FFTXlib and LAXlib) 
that originated from Quantum ESPRESSO suite and tailored for plane wave basis, 
will be inserted in the ESL bundle in future releases.


\subsection{\wannier}\label{sec:wannier}

Wannier functions (WFs)~\cite{wannier-pr37,MMYSV_RMP84} provide a localised real-space representation of the electronic structure of materials that is complementary to the reciprocal-space representation of Bloch bands.
The freedom associated with the choice of gauge of Bloch states can be used to construct exponentially localized WFs.~\cite{Brouder_PRL_2007} So-called \textit{maximally-localized Wannier functions} (MLWFs)~\cite{MV_PRB56} are obtained by choosing the gauge that minimizes the total quadratic spread of the WFs. 
Both the case of isolated bands,~\cite{MV_PRB56} i.e., a composite set of bands that is separated from other bands in the Brillouin zone (BZ) by energy gaps, and the case of entangled bands~\cite{SMV_PRB65} can be treated.

MLWFs are used routinely, for example, to analyse and understand chemical bonding, to perform high-accuracy fine-grained interpolation of quantities in the BZ (such as band energies, Berry phase properties and electron-phonon interactions), to characterise topological materials, to construct compact tight-binding models of materials, and to compute charge transport properties. We direct the reader to Refs.~\onlinecite{MMYSV_RMP84} and \onlinecite{Pizzi-2020} for details of the underlying theory of MLWFs and their diverse applications. Instead, here we focus on (1) the aspects of the \wannier\ code~\cite{Pizzi-2020, wannier-website, wannier-github} and its development that have enabled it to emerge as a paradigmatic example of an interoperable software tool, and (2) future plans that will take the code further in directions that reflect the broader philosophy of the \esl. 

From its conception,~\cite{Mostofi_CPC} \wannier\ was designed to make the addition of new functionality as easy as possible, by being modular, well-documented and well-commented, and to be as independent as possible from the underlying code that calculates the Bloch bands from which the MLWFs are constructed. As such, \wannier\ requires only the matrix elements 
$M_{mn}^{(\mathbf{k,b})} = \left\langle u_{m\mathbf{k}} | u_{n\mathbf{k+b}} \right\rangle$,
where $u_{n\mathbf{k}}(\mathbf{r})$ is the cell-periodic part of the Bloch function $\psi_{n\mathbf{k}}(\mathbf{r})=u_{n\mathbf{k}}(\mathbf{r})\mathrm{e}^{i\mathbf{k}\cdot\mathbf{r}}$, together with an initial guess for the choice of gauge. The latter can be obtained either by projecting an appropriate set of atomic-like orbitals $g_n(\mathbf{r})$ onto the initial Bloch states, or by using the recently implemented ``selected-columns of the density matrix'' (SCDM) method,~\cite{Vitale2019arXiv,DL_2015_SCDM,DL_2018_SIAM} that does not demand the human intervention often needed to define good projection functions.

Since these matrix elements, together with the eigenvalues of the single-particle hamiltonian, are independent of the specific implementational details of the underlying electronic structure code (e.g., basis set, grids, symmetry operations, level of theory, pseudopotentials), \wannier\ is fully interoperable with any code that is able to calculate them. The onus is largely on the developers of electronic structure codes, therefore, to develop and maintain their own interface that provides these quantities. \wannier\ allows for two interface modes: (1) via reading and writing files to/from disk and running \wannier\ as a separate external executable; and (2) via calls to the \wannier\ library directly from within a program. Electronic structure codes that interface to \wannier\ include: Quantum ESPRESSO,~\cite{giannozzi_qe_2017} ABINIT,~\cite{Gonze2016_106} VASP,~\cite{kresse_cms_1996} \siesta,~\cite{siesta_soler_2002} WIEN2k,~\cite{Blaha2018} Fleur,~\cite{blugel_fleur_2006}, \octopus,~\cite{andrade2012time,octopus_2020} ELK,~\cite{elk} \bigdft,~\cite{Genovese2008_014109} GPAW,~\cite{mortensen_gpaw_2005} pyscf,~\cite{PYSCF} and openmx.~\cite{openmx-w90_PRB2009} New developments in \wannier, therefore, are available to the vast majority of the user community rapidly, which serves to accelerate research. 

The most recent major release (v3.x) of \wannier\ is able to compute a growing range of properties,~\cite{Pizzi-2020} a range that is increasingly difficult to maintain in one code with a small group of developers. Furthermore, there is a growing community of researchers and codes, including Gollum,~\cite{ferrer_gollum_2014} WannierTools,~\cite{wu_wanniertools_2018} NanoTCAD ViDES,~\cite{nanotcadvides} Yambo,~\cite{marini_yambo_2009} Z2Pack,~\cite{gresch_z2pack_2017} Triqs,~\cite{parcollet_triqs_2015} and EPW~\cite{ponce_EPW_2016} that use \wannier\ to calculate an even wider range of properties. For these reasons, in 2016, ten years after its first release,~\cite{Mostofi_CPC} \wannier\ transitioned to a community-development model in which the code is hosted on GitHub~\cite{wannier-github} and community-driven developments are invited via a fork and pull-request approach. Code integrity is maintained via a documented coding style guide that contributors must adhere to, together with nightly automated building and testing on a Buildbot~\cite{buildbot} test farm, and continuous integration with Travis CI~\cite{travis}, whereby a pull-request triggers a suite of test calculations and is blocked if any tests fail.

What does the future hold for \wannier? Currently, only a small subset of the full functionality of the code is accessible in the library mode of \wannier, and only in serial processing. The next major planned development, therefore, is to completely re-engineer the library mode of the code such that the full functionality of \wannier\ (including parallel processing) is accessible via library calls from within, e.g., an overarching workflow, a dynamical simulation, or a self-consistent field iteration. This would enable advanced materials properties to be calculated seamlessly \textit{on the fly}. Using the code in this way is made significantly more practical due to recent developments in generating MLWFs automatically with minimal user-intervention.~\cite{Vitale2019arXiv} Some challenges will need to be overcome, however, including: determining the optimal strategy for parallelisation given that this is likely to conflict with that of the host electronic structure code; and handling errors thread-safely yet unobtrusively. When this development is complete, like the ouroboros that swallows its own tail~\cite{ouroboros-wiki}, we envisage the main \wannier\ code becoming a wrapper for its own library calls. 

\subsection{\matrixswitch}
\label{matrixswitch}

The \matrixswitch library was developed alongside \libomm during the first \esl workshop, but is independent of it. Its aim is to act as an intermediary layer between high-level routines for physics-related algorithms and low-level routines dealing with matrix storage and manipulation, allowing the former to be written in a way which is close to mathematical notation, while also enabling seamless switching between different matrix storage formats and implementations of the matrix operations. As new formats are introduced in \matrixswitch, they can immediately be used in the high-level routines without any further modification of the code. Both dense and sparse formats are supported, as well as serial and parallel distributions.

At the centre of \matrixswitch is the \texttt{matrix} object, a Fortran derived type defined by the library which acts as a wrapper for the specific storage format. A small number of basic operations are defined for this object, such as setting and getting elements, matrix-matrix multiplication, matrix addition, traces, etc. The API is also easily extensible to include more complex matrix operations which are not part of the standard set, as the object is quite transparent and can be unpacked when needed to operate directly on the underlying data structures.

An additional feature of the library is that it facilitates its usage only for a subset of the host code (e.g., in a specific module), by permitting pre-existing matrix data conforming to one of the supported storage formats to be simply registered by \matrixswitch without the need for copying, converting, or allocating new memory. The registered matrix can then be used for any \matrixswitch operation as if it were natively managed.

As well as being used by \libomm, \matrixswitch is also currently being used in parts of \siesta (e.g., for the recently developed real-time time-dependent DFT algorithm) and in smaller codes used for individual research projects.~\cite{Corsetti_2017}
  \matrixswitch has recently been extended\cite{dbcsr-matrix} to use the DBCSR library~\cite{DBCSR-2014,DBCSR-site} 
(Distributed Block Compressed Sparse Row), a linear-algebra parallel engine for sparse matrices.
  The original \siesta linear-scaling solver based on OMM with finite support
  solutions~\cite{Ordejon-1995} is being refactored to use \matrixswitch
  and thus take advantage of the DBCSR backend. This strategy will be
  extended to other related solvers.~\cite{linear_bowler_2012}

\subsection{\flook}

  The \flook library was developed with the objective to control flow and move 
lightweight operations into scriptable code.
  It provides a simple way to code top steering tools (see Fig.~\ref{fig:models} b)
such as molecular dynamics (MD) methods which are called only once every 
MD step and are typically light operations, but it also allows deeper-level 
control of the code, such as the tuning of mixing parameters or stopping 
calculations when certain criteria are met, for example.

  The scripting can be efficiently implemented by embedding a Lua
interpreter into the application program.
  Lua is a lightweight embeddable scripting language.\cite{luabook}
The software library \flook enables Fortran and Lua to communicate together in a seamless way by passing variables to and from ``tables'' in Lua. 
Having a hook between Lua and Fortran empowers end-users to create their own scripts in Lua in order to extend the functionality of codes. Since any data can be moved between Fortran and Lua, the Lua script which implements a particular functionality can be used to replace that functionality inside the core Fortran code, if so desired.

This methodology works by assigning Lua functions to be called in Fortran. By populating a table with the internal data-structure in Fortran using dictionaries one can easily enable variable passing between Fortran and Lua using a single line of code. As an example, here is a snippet of Fortran and Lua code which enables access to the atomic coordinates in the \siesta  DFT package.

\paragraph{\textbf{Fortran code:}}
\begin{verbatim}
type(dictionary_t) :: variables
! Add atomic coordinates to table of variables
variables = variables // ('geom.xa'.kvp.xa)
\end{verbatim}
\paragraph{\textbf{Lua code:}}
\begin{verbatim}
-- Retrieve atomic coordinates and manipulate
local xa = siesta.geom.xa
\end{verbatim}

Currently \flook is used in \siesta to extend molecular dynamics methods, customize outputs, force-constants calculations and convergence of precision parameters. It also exposes convergence variables which allows the user to change these  parameters while a calculation is running. A derived project (flos [\url{https://github.com/siesta-project/flos}]) implements the scriptable Lua functions that may be used for other projects using \flook.

\subsection{\libfdf}

FDF stands for Flexible Data Format, designed within the \siesta
project to simplify the handling of input options. It is based on
a keyword/value paradigm (including physical units when relevant), 
and is supplemented by a block interface for arbitrarily complex 
blobs of data.

\libfdf\cite{weblibfdf} is the official implementation of the 
FDF specifications for use in client codes.
At present the FDF format is used extensively by \siesta, and it has
been an inspiration for several other code-specific input formats.

New input options can be implemented very easily. When a keyword is not present
in the FDF file the corresponding program variable is assigned a
pre-programmed default value. This enables programmers of client codes
to insert new input statements anywhere in the code, without worrying 
about ``reserving a slot'' in a possibly already crowded
fixed-format input file.

\subsection{xmlf90}

  xmlf90 is a package to handle XML in Fortran. 
  It has two major components: 
  ($i$) A XML parsing library, with the most complete programming 
interface based on the very successful SAX (Simple API for XML) model,\cite{SAX}
although a partial DOM interface and a very experimental XPATH interface
are also present.
  The SAX parser in particular was designed to be a useful
tool in the extraction and analysis of data in the context of
scientific computing, and thus the priorities were efficiency and the
ability to deal with potentially large XML files while maintaining a small
memory footprint. 
  ($ii$) A library (xmlf90-wxml) that facilitates the writing of well-formed
XML, including such features as automatic start-tag completion,
attribute pretty-printing, and element indentation.
  There are also helper routines to handle the output of numerical 
arrays.
  xmlf90 is the parsing engine for the \libpsml library of 
Section~\ref{sec:psml}.

\section{The \eslbundle}\label{sec:bundle}

\begin{figure}[!htb]
    \centering
    \includegraphics[width=0.95\columnwidth]{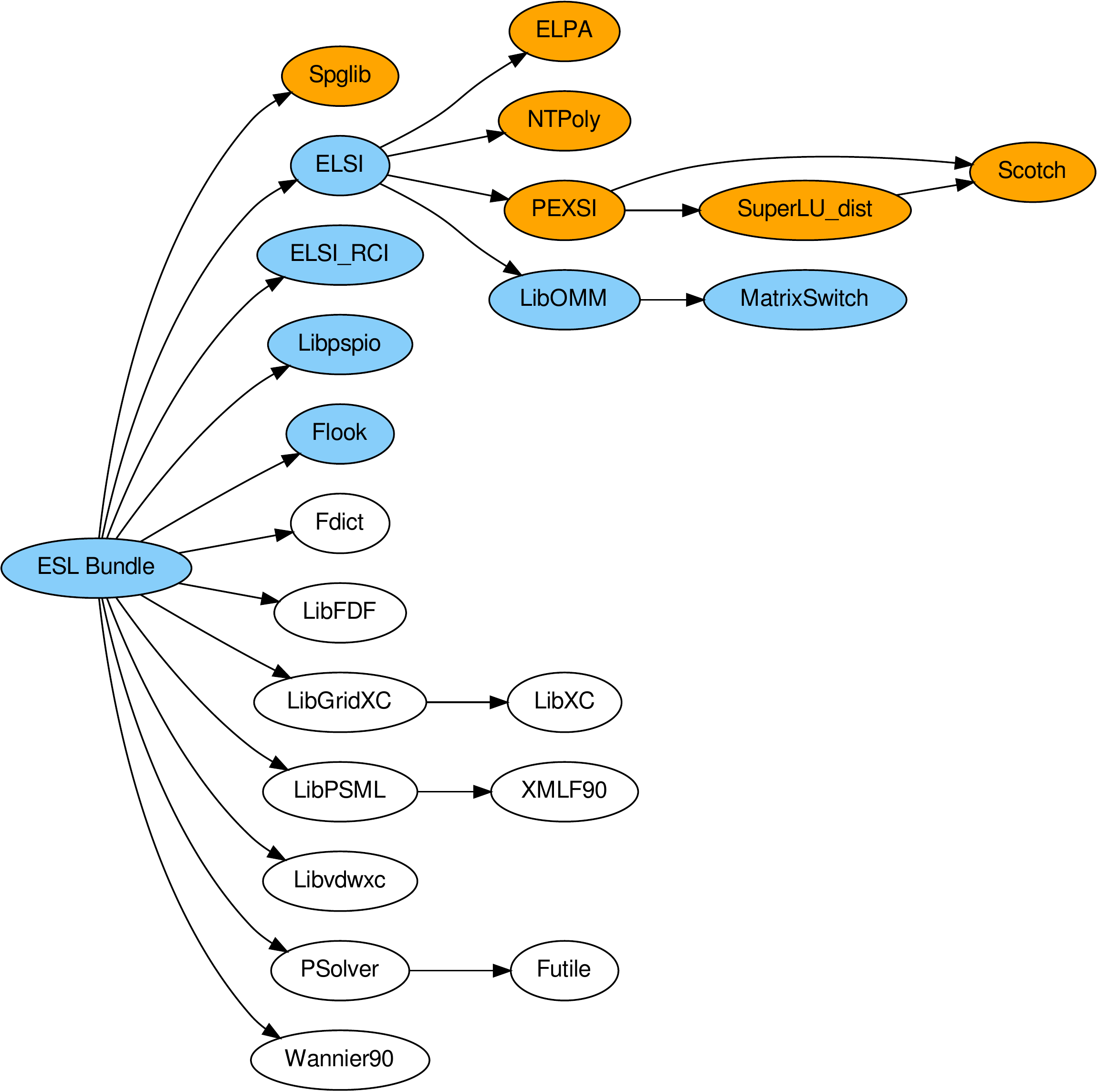}
    \caption{Internal dependencies between the components of the \esl Bundle. White background: components extracted from electronic-structure codes participating in the ESL. Blue background: components created within the ESL or through its activities. Orange background: components maintained outside the ESL. The complete tree of solvers accessed by \elsi appears in
    Fig.~\ref{fig:elsi}.}
    \label{fig:esl-bundle-deps}
\end{figure}

Adopting a modular approach has many advantages: smaller software units to develop and maintain,
easier testing of each component, faster propagation of fixes, better separation of technical
domains, reduced duplication of code, to cite a few obvious ones. However, it also implies some
risks: if not addressed explicitly, the asynchronous evolution of the individual components --- 
aka modules --- quickly becomes a severe obstacle to the improvement and maintenance of the whole.
\libxc is an emblematic example of this kind of situation: whenever a new exchange-correlation
functional is implemented in \libxc, a few dozen DFT codes are just one compilation away from using
it. However, if the API of \libxc changes, each DFT code has to update its interface to \libxc to
be able to use the new version, which results in a situation where some codes use one version of
the library while others are stuck with an older version.

When one module depends on another, or a number of others, a few aspects have to be considered with great care:
\begin{itemize}
    \item A given version of the dependent module is compatible with only some versions of those it is dependent on.
    \item Even when the two versions are compatible, not all configuration options available will actually work, i.e. the two modules may have conflicting requirements in some situations.
    \item In addition to technical aspects, social considerations have to be taken into account, in particular when the two modules are developed by different teams.
\end{itemize}

To mitigate these risks, we provide all the \esl software libraries in the form of a bundle. The \eslbundle provides a set of ready-to-use software modules
such that to each version of the \eslbundle there corresponds a well defined set of module versions that are compatible among themselves.
The contents of the \eslbundle are curated through the activities of the \esl
and supervised by the \esl Steering Committee (for more details about the governing structure of the \esl, see Appendix~\ref{sec:steering}). 
Adding, updating or removing modules is discussed during \esl workshops and Steering Committee meetings until an agreement is reached, 
before being thoroughly tested to detect possible compatibility issues. The validation of any change within the \eslbundle{} sometimes 
involves a high level of complexity, which is why it is performed by a team of volunteers and includes manual steps. 
Indeed, the \eslbundle{} is meant to be used in production by ESC codes, not just to be successfully installed on a given set of systems. 
What will finally decide whether a module can be updated will be the usability of its new version by these ESC codes, 
which is why complementary tests should be conducted with the codes themselves before releasing a new version of the \eslbundle{}.
Fig.~\ref{fig:esl-bundle-deps} summarises the dependencies between individual modules, 
which are actively monitored in collaboration with their respective developers. 
Changes and improvements brought by the \esl are 
reported to the original developers of the affected modules and contributed back to the upstream module whenever possible.

However, providing a bundle by itself is not enough. To ensure its usability, the \eslbundle{} must be easy to compile and install on different platforms and by users with different needs and goals. 
With so many different components, written in different programming languages and using different build systems, this is far from being a trivial task.
This is why the \eslbundle needs to be distributed in different forms, each targeting a different use case. We describe two of these distribution channels in greater detail in the following subsections. 
A third distribution channel currently under consideration is to provide the \eslbundle as Debian and RPM packages. Several components, like \libxc, are already
included in the official repositories of several popular Linux distributions, like Debian or Fedora,
as well as in the MacPorts package manager for macOS,
but we would like to extend this to the whole \eslbundle. A fourth channel of distribution for the bundle is the collection of docker images released publicly on docker hub, at \url{https://hub.docker.com/u/eslib} . After each release a new docker image is built using JHBuild scripts, tagged and used to test the \esldemo. These images can be handy for quick access to a binary distribution of the bundle, of benefit to both developers and curious users.

\subsection{JHBuild bundler}

To provide the \eslbundle in a fully self-contained way with a common installation interface for all of its components, we use the JHBuild framework.~\cite{jhbuild}
JHBuild is an actively-maintained Python build framework used by the GNOME Project\cite{gnome}, an open-source desktop environment for Unix-like operating systems, 
which has been solving the same challenges as the \esl over the last two decades. JHBuild is able to build a collection of modules, that it names \textit{modulesets}, 
from a minimal amount of information: download URL, type of build system, and one-to-one dependencies, plus optional on-the-fly corrections (patches). 
JHBuild determines the correct order of compilation of the modules by itself and strictly separates the aspects related to the modulesets from those belonging to the build environment. 
The latter is achieved by using configuration files to tune the build parameters, globally or for each module.
In the case of the \eslbundle, we provide a curated collection of such configuration files for Linux-based systems and macOS. 
The use of JHBuild greatly simplifies the installation of the \eslbundle and is ideal for developers or users of electronic structure codes 
that require one or more components to be installed on their personal computers, but do not care too much about performance.

\subsection{HPC-oriented distribution}

Once we consider software provisioning for HPC resources, where software such as the \eslbundle should leverage the available hardware and seamlessly integrate into the existing software stack,
the situation becomes vastly more challenging. In this context, the ESL is far from alone in the depth and complexity of its software stack. Application developers, HPC sites, and end users around
the world spend significant amounts of time creating and verifying optimised software installations for such resources. Although the problems that arise with installing scientific software are ubiquitous,
there is currently inadequate collaboration between HPC sites and/or HPC domains. At the \emph{``Getting Scientific Software Installed''} Birds-of-a-Feather session at SC'19 less than 30\% of the survey 
respondents answered 'yes' when asked whether they work together with other HPC sites on software installation.

EasyBuild~\cite{easybuild} is a tool for providing optimised, reproducible, multi-platform scientific software installations in a consistent, efficient, and user-friendly manner. EasyBuild is currently 
used by well over 100 HPC sites worldwide (including J\"ulich Supercomputing Centre, CSCS, Compute Canada, SURFsara, SNIC, \ldots). Leveraging EasyBuild for HPC-oriented distribution provides 
the \esl with an HPC-oriented build infrastructure that can quickly and reliably distribute the bundle to a large number of HPC sites.

EasyBuild employs so-called compiler toolchains, or simply \emph{toolchains} for short, which are a major facilitator in handling the build and installation processes. 
A typical toolchain consists of one or more compilers, usually put together with some libraries which provide specific functionality, e.g., for using an MPI stack for distributed computing, 
or which provide optimized routines for commonly used math operations, e.g., the well-known BLAS/LAPACK APIs for linear algebra routines. For each software package being built, 
the toolchain to be used must be specified. Notably, EasyBuild already supports over 1800 software packages, including many of the (direct and indirect) dependencies of 
the \eslbundle{}. These verified and consistent infrastructures allow \esl development efforts to focus primarily on its component libraries which can be synchronised with the EasyBuild toolchain
release cycle (which currently has two updates per year). This is why EasyBuild replaces JHBuild for the distribution of the \eslbundle{} on HPC environments.

\section{Use cases in end user codes}

\begin{figure*}[t] 
\centering
\includegraphics[width=0.9\textwidth]{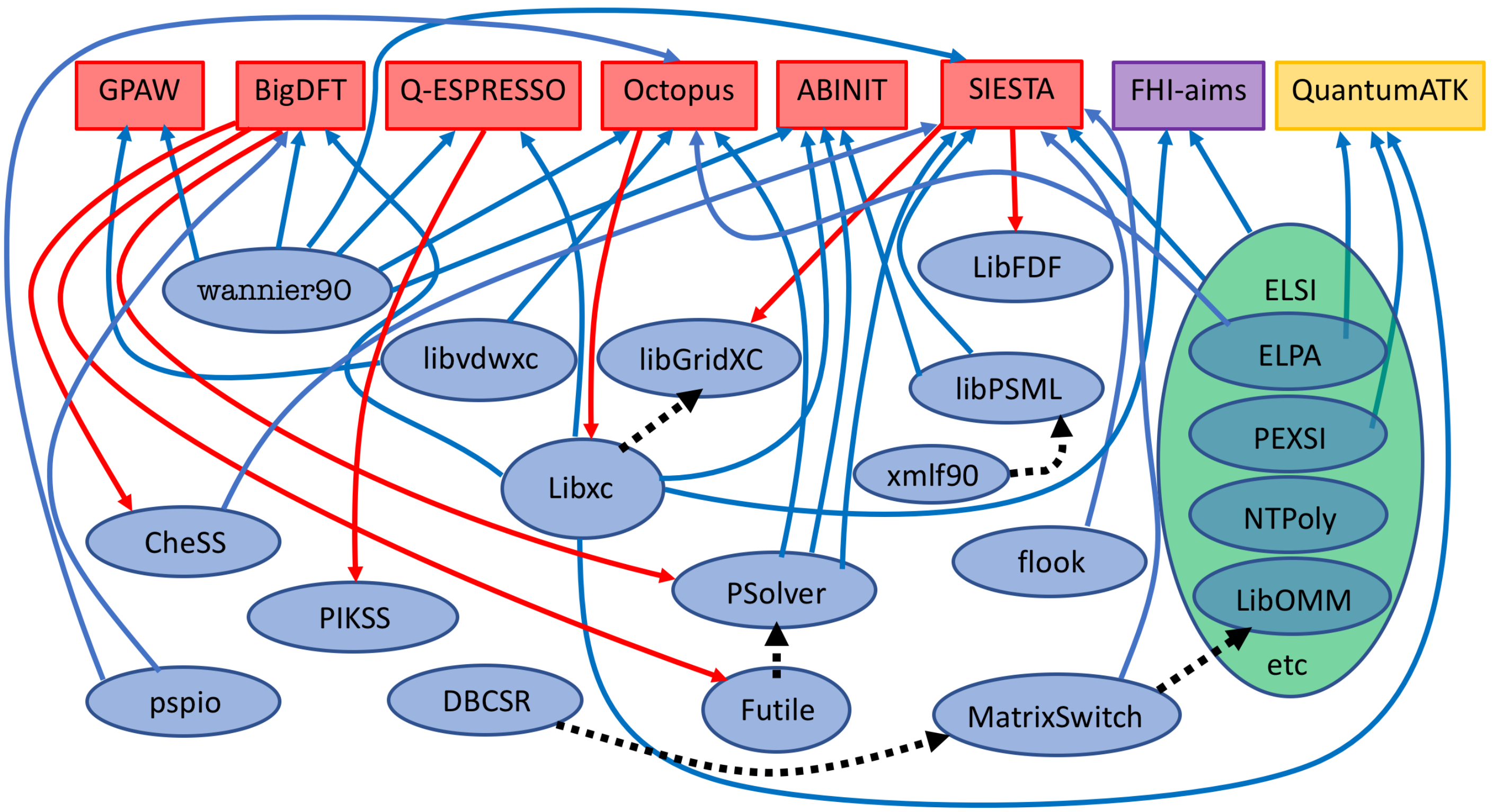}
\caption{Use of \esl libraries within electronic structure codes.
The rectangular boxes at the top show electronic structure programs
(end users), red indicating open-source programs (licensed via GPL),
while \fhiaims is distributed by a registered non-profit organization (Molecular Simulations from First Principles e.V., MS1P -- https://ms1p.org/) under a proprietary licence, and QuantumATK is under 
commercial licence distributed by the software company Synopsys. 
  This set contains codes connected to the \esl, mostly via 
contributors to \esl being developers of these codes (there are
many other codes that use at least some of the depicted libraries).
  Blue ellipses indicate libraries described in this
paper. 
  The larger (green) ellipse corresponds to \elsi as a general
interface to several Hamiltonian solvers, with its associated
libraries.
  Arrows indicate dependencies. Thick black dashed lines indicate libraries
dependent on other libraries, blue lines show libraries directly used by the 
codes, and red lines indicate libraries that were re-engineered by
extracting them from a particular code (and are also used by that code). 
DBCSR~\cite{DBCSR-2014,DBCSR-site} is included here because it has 
been coupled~\cite{dbcsr-matrix} to \matrixswitch as a parallel 
sparse linear-algebra engine. 
  \elpa is also used by \abinit, QuantumESPRESSO, and GPAW.}
\label{fig:usage}
\end{figure*}

  Fig.~\ref{fig:usage} illustrates the usage of \esl packages by
the electronic structure programs engaged in the \esl project.
  There are more cases of usage not covered here, namely, other 
electronic structure codes which use some of the libraries described above.
  As stated previously, \esl collects both libraries that have been
built or extracted from codes purposely for \esl, together with
independently developed and maintained libraries, as e.g., 
\libxc and \wannier, which predate \esl, and whose authors
agree with (and contribute to) their incorporation into
the \esl. 
  The libraries in Fig.~\ref{fig:usage} are the ones that are 
(or are being) included in the \eslbundle described in 
Section~\ref{sec:bundle}. 

  The lines in Fig.~\ref{fig:usage} show dependencies between components 
of the \esl and user codes. 
  The red lines indicate dependencies on libraries (depicted as ovals) 
that have been extracted as independent library components of the \esl 
from the connected user code (the rectangles).
  Of course, the original codes use them as well.
  The blue lines show which user codes use components that were independently
developed in the \esl. 
  Some of the libraries, although not extracted from any given code, were
developed by code developers of a particular code.
  However, such connections are not indicated in the figure.
  In the following we describe the links and \esl usage illustrated
in Fig.~\ref{fig:usage} for the codes shown, starting with
the \esldemo, a very lean electronic structure code created
from scratch in a couple of weeks and built on the \esl.

\subsection{\esldemo}
\label{sec:esl-demo}

Four years after the \esl was initiated, the \esl team realised it had gathered sufficient libraries to account for nearly 
all the complex parts of a simple DFT code. In February 2018, the 5th \esl workshop was focused on building an entire DFT code 
from scratch, within a fortnight.
The purpose of such demonstrator code is to showcase the usage of \esl libraries and to provide
a framework to test the \eslbundle. It is not in any way intended to be a competitor
to existing DFT codes. Instead, it can be seen as being part of the \esl documentation,
guiding new users and developers of the \esl. As such, some effort has been made to make it clear, simple and easily extendable.

\begin{figure}[!htb]
\centering
\includegraphics[width=0.4\textwidth]{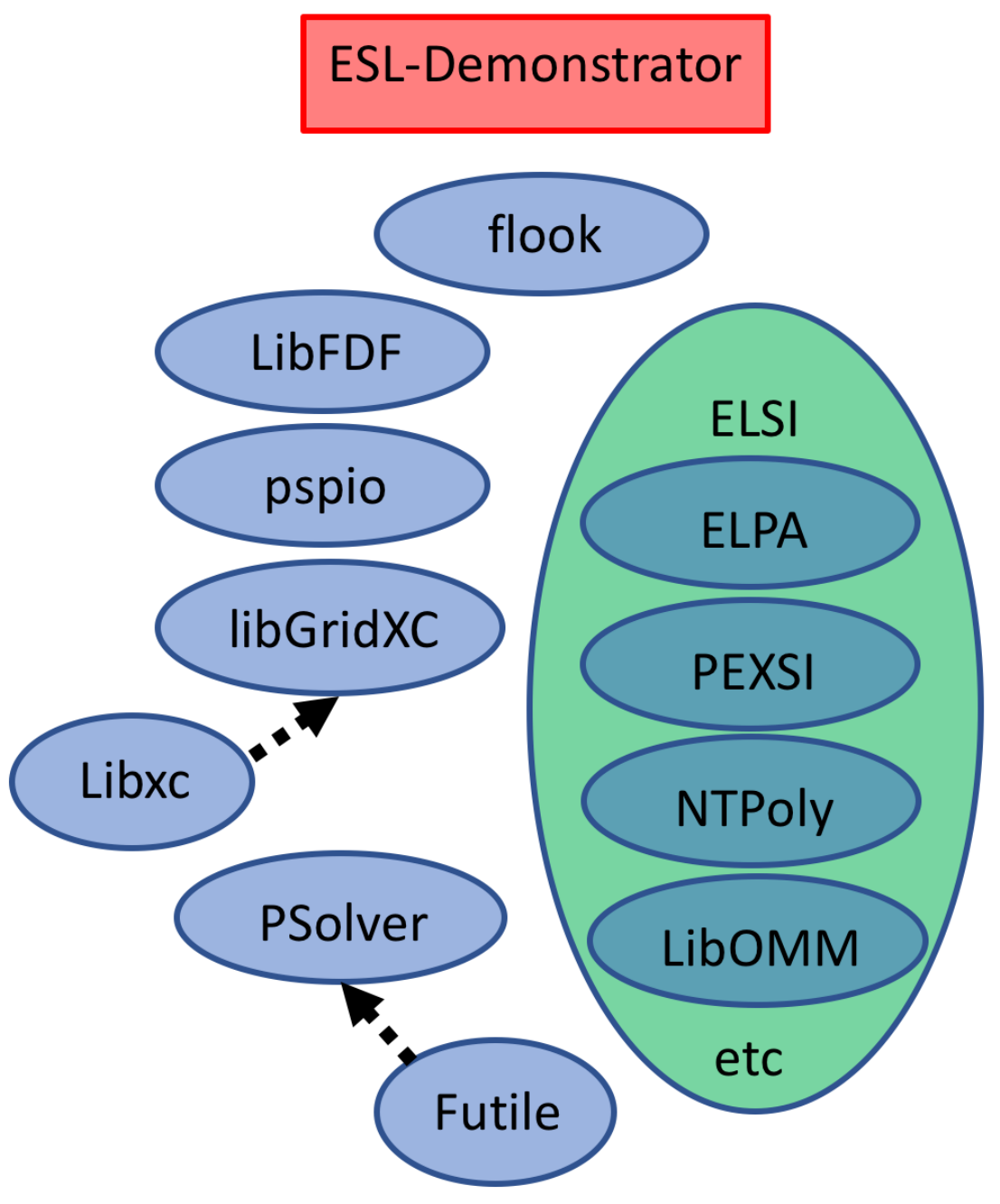}
\caption{Libraries used in the full DFT 
program developed as a demonstrator for the \esl. 
  It allows the user to choose between plane-waves or atomic orbitals
as basis sets.
  The libraries themselves are as in Fig.~\ref{fig:usage}, and 
as described in this section.}
\label{fig:demo}
\end{figure}

The resulting code, the \esldemo, is a functional DFT code which makes extensive use of the \esl libraries
presented in Fig.~\ref{fig:demo}.
  It uses \pspio for reading pseudo potentials,
\psolver to calculate the Hartree potential, \libfdf as the input engine,
\libgridxc to calculate the XC potential on a grid,
\libxc to evaluate the XC functional on the points of that grid,
\elsi for calculating eigenstates, and \flook to make scriptable 
control flows.
  During the development it was also decided to follow the Sphinx documentation 
style to retain a unified documentation scheme.

The development of the demonstrator code was carefully divided between teams formed from the 14 people who attended the workshop coding session.
The tasks were assigned to suit the individual expertise within each team whilst also taking into account the backbone code of the demonstrator. 
In particular there are several parts of a DFT program required, 1) user input, 2) Hartree potential, 3) XC potential, 4) eigenstate solver, 4) scriptable work-flows.
While most codes use either a plane-wave or a localized orbital basis, the \esldemo allows users to use either of the two.
Such a decision makes the code slightly more complicated, but it allows an increase in the range of libraries used and provides
newcomers to DFT codes an easy access point to two classes of basis sets that require very different numerical methods.

The \esldemo successfully uses the aforementioned libraries. 
It currently only allows non spin-polarized, $\Gamma$-point calculations 
as well as serial execution. 
Some of the missing features are due to shortcomings in the \eslbundle. For example, it currently contains no 
libraries to generate $\mathbf k$-point grids, which restricts the \esldemo to $\Gamma$-point only calculations.
Other missing features are not due to the underlying libraries, but simply reflect the early stage
of development of the \esldemo. Our plan is to keep extending the \esldemo to cover more features provided by existing \esl libraries and by new libraries added
to \eslbundle in the meantime. Work is underway on allowing parallel execution and spin-polarized calculations.

An important effect of developing the \esldemo is the exposure of 
possible bugs, missing features and testing how integrable and inter-operable libraries actually are.
  Indeed the development of the demonstrator led to the discovery of 
certain bugs and build problems in \esl libraries. The \esldemo acts a \emph{de facto} test for the \eslbundle where a successful build and run of the demo is a prerequisite to release a new bundle. 
  The code is hosted here \url{https://gitlab.com/ElectronicStructureLibrary/esl-demo}.

\subsection{\esl in participating codes}

\subsubsection{\abinit}

   \abinit pioneered the open-source model within the electronic-structure
community.
   It is a plane-wave-based code that has allowed contributions from quite 
an open community of developers, some of them coding new features, tools, 
etc. in modules being integrated into the code.
  In that sense it can claim to have taken the first intellectual 
step within the electronic-structure community that led to the \esl.
  In addition to this contribution to the \esl concept, \abinit  
benefits from the usage of \libxc for the local and semi-local 
exchange-correlation terms, and it now incorporates \libgridxc 
on top of \libxc for the global grid treatment on non-local functionals. 
  It also makes use of \psolver for the computation of the Hartree terms 
to energy and potential, and the PSML standard and associated library 
for pseudopotential input.

\subsubsection{\bigdft}
\label{sec:bigdft}

For several years already, the monolithic sources of \bigdft have been divided into several subdirectories, that slowly became independent from each other and were finally separated into their own modules, living in a separate Git tree and that are shipped with their own build system. This direction of work was seen by the developer team as a way to keep the development sustainable in terms of functionalities and maintenance. It started with a library implementing a tool box for Fortran, Futile that is available in the \esl. This tool box started with in-memory representation of a YAML document,\cite{yaml} but was quickly extended to keep track of dynamic memory allocations, time measurements, error handling, etc.
  The versatile Poisson solver used in \bigdft is now completely independent 
from its origins and also available in the \esl.
  Some other components of \bigdft available in the \esl, were developed 
from the start as separated libraries, like the sparse matrix library CheSS.\cite{chess}
  \bigdft is also taking advantage of codes developed outside the project.
  Like numerous other DFT codes, it is using \libxc for the exchange and
correlation calculation.
  An interface exists to post-process the calculated wave functions using 
\wannier.
  While initially coded for the Hutter-Goedecker-Hartwigsen pseudopotential
formalism,\cite{Hartwigsen1998} a link with \pspio was written to allow 
a greater range of pseudopotentials.
  Finally, \bigdft is distributed as a bundle, like the \esl. 
  It takes care of the compilation and linking of the various libraries 
and the end project itself, to deliver a single executable to the end user.

\subsubsection{\fhiaims}


  \fhiaims developers have greatly contributed to the \esl through joint involvement in
the broader, U.S. NSF-funded ELSI project, described in Section~\ref{sec:elsi}. \elsi was inspired by the \esl and represented an early 
U.S. initiative in this effort, in an otherwise primarily
European endeavour.
  Through \elsi, \fhiaims now benefits from a range of 
Hamiltonian solvers in a seamless framework. This list includes \elpa, which originated within \fhiaims and is maintained as a standalone solver library led by the Max-Planck Computing and Data Facility, as well as \pexsi, \ntpoly and all other solvers supported by \elsi.
  The libraries used by \fhiaims are not restricted to solvers,
as \libxc is supported for the calculation of the exchange-correlation 
contribution for local and semi-local functionals.

\subsubsection{GPAW}

  The GPAW code allows for different modes of operation, according
to the way Kohn-Sham wavefunctions are represented.\cite{mortensen_gpaw_2005,Larsen2009}
  They are based on, namely, finite differences (FD), plane waves (PW), 
or linear combination of atomic orbitals (LCAO).
  The LCAO mode uses \elpa for fast parallel diagonalization of the 
Hamiltonian matrix.
  GPAW also uses \libxc plus \libvdwxc to support LDA, GGA, 
meta-GGA, and vdW-DF exchange-correlation functionals.

\subsubsection{Multiple scattering codes}

  Codes built around the computation of the Kohn-Sham Green’s 
function, by means of multiple scattering (MS) theory, give immediate 
access to spectroscopic properties, transport and many other response 
functions.
  In addition, these methods can deal with many different problems in 
electronic structure theory for systems with and without periodicity, 
such as disordered alloys and semi-infinite surfaces.\cite{Ebert2011} 
  Multiple-scattering codes typically import data such as self-consistent
Kohn-Sham potentials or charge densities from other ES codes. 
  The set of ESCDF format specifications and the associated library 
is ideal for that purpose. 
  It is being already used by data transfers between codes 
such as the Munich SPR-KKR\cite{Ebert2011,Ebert2016} and
MSSpec.\cite{Sebilleau2011}

\subsubsection{\octopus}

\octopus is currently interfaced to several libraries that are part of the \esl Bundle. 
  The \libxc library, although it is now completely 
independent, was originally developed within \octopus.
When treating finite systems in \octopus, the default method to solve Poisson's equation is the one provided by \psolver.
Evaluation of exchange and correlation functionals that depend explicitly on the density is done exclusively using the \libxc and \libvdwxc libraries.
Support for reading pseudopotentials using the \pspio library is also provided, as well as the possibility of using \wannier\ to compute maximally-localised Wannier functions 
from the Bloch states. Finally, the \elpa library can be used whenever direct diagonalization of matrices is required.

\subsubsection{QuantumATK}

QuantumATK is a commercially-developed platform which includes its own LCAO and plane-wave DFT solvers, as well as semi-empirical tight-binding and force-fields. The code is closed-source, but makes use of several external software libraries; among these are three libraries in the \eslbundle: \libxc, \elpa and \pexsi (the last two included independently of \elsi). \elpa and \pexsi are used not only for the LCAO-DFT solver but also for the various semi-empirical tight-binding solvers.

QuantumATK is a unique case amongst the list of current codes using \esl libraries, for a number of reasons: (a) its closed-source and commercial nature means that there are strict constraints on the licensing of libraries it can use (the most common being MIT, BSD and LGPL); (b) it uses C++ as its backend language (with a Python frontend), and the \esl libraries are therefore linked to C++ rather than Fortran or C; and (c) executables are compiled and shipped for Windows as well as Linux.

\subsubsection{Quantum ESPRESSO}
\label{sec:QE}

  This plane-wave program distribution contains a variety of optimised
iterative Hamiltonian solvers tailored for the plane-wave basis. 
  Within the \esl effort, they were extracted and isolated into the PIKKS
KS-Solvers suite and library, together with additional components
to perform fast-Fourier transforms (FFTXlib) and parallel linear
algebra operations (LAXlib). These will be inserted in the ESL bundle in future
releases.
  Use of these components is demonstrated in a simple empirical
pseuodopotential code that can be used as tool for further 
developments.\cite{PIKSS}
  Quantum ESPRESSO codes link to these libraries, as well as to 
other \esl libraries of different origin, such as \libxc, \elpa, and 
\wannier.

\subsubsection{SIESTA}

Two libraries now in the \esl (\libgridxc, \libfdf) originated as modules
within \siesta. Several more (\flook, xmlf90 and \libpsml) were developed with
general usefulness in mind but also to address issues of relevance to that
program. \siesta is thus an important contributor to the \esl. In the
opposite direction, \siesta benefits from other \esl-provided
functionality, most clearly in the area of solvers, with an interface
to \elsi that has significantly extended the choices available and
enhanced the performance of the code. The \libxc library is also used
through the interface to \libgridxc. 
  \wannier\ has also been fully incorporated as a library.
  Work is now being done to
incorporate the new functionality available in the \psolver
library in \siesta and there are plans to benefit from some of the low-level
utilities in the Futile package.

\section{Future} 

  The \esl represents a channel for possible spontaneous
utility projects to develop and link into present and new 
electronic structure packages.
  In this sense, the future evolution of the \esl from the
point of view of the sub-packages it contains is quite 
open.

  For the mid- and long-term future of the \esl, a key metric of success will
be wide usage.
  In addition to communication (as done in this paper and on
the web), the following aspects will be important. 
  ($i$) Content -- useful features. High-level programmers should be
able to find in the \esl key tools for their programs.
  ($ii$) Performance. The libraries will need to be maintained,
keeping up with hardware evolution, and maintaining competitive
standards of efficiency and scalability.  
  An important component of future performance will be the
definition and stabilisation of APIs, in addition to good
standards of documentation.
  ($iii$) Easy use. The library has to be user friendly, not necessarily
for the end user, but for computational-scientist coders, who
will implement new codes and/or features linking to the ESL.
  ($iv$) Easy build. End users of programs that link to the \esl
should be able to compile their codes reasonably easily.
  
  Concerning content, there is a list of candidate libraries to
include, as well as modules in present programs that can be 
extracted as libraries.
  In the short term, there are packages (some of them mentioned above)
that are being prepared for inclusion into the ESL and its bundle.
  This is the case for the CheSS library,\cite{chess} which implements
a linear-scaling Hamiltonian solver based on a Fermi-operator expansion.
  It arises from the \bigdft program, but it already works as 
a separate library, and is already used by other codes, such as
\siesta.
  Similarly, the connection between \matrixswitch and DBCSR,
illustrated in Fig.~\ref{fig:usage} will soon be bundled into the
\esl.
  Also a candidate for bundling into the \esl is the libPAW library,\cite{libPAW} 
currently distributed in the \abinit package, but also used in \bigdft 
and other codes.
  It is a collection of objects and routines intended to facilitate 
the porting of the projector augmented-wave method ``out of the box'' 
onto any ES code regardless of the basis used for the wave functions.
  DFTB+\cite{dftb-plus-new,dftb_aradi_2007} developers are also joining the 
\esl effort contributing their semi-empirical electronic structure engine 
and the stand-alone SAYDX library (Structured Array Data 
Exchange),\cite{saydx} which is an auxiliary library 
that provides a platform for 
exchanging array data using a simple tree structure.
  It offers a framework to build, manipulate and query such 
array data trees, as well as send and receive them through various 
transport layers.
  They will be incorporated into future \esl bundles.

  Although the \esl and this paper focus on electrons, 
an important line of future work is the incorporation of
upper-level steering packages and libraries, prominently
molecular-dynamics engines, and, more generally, codes dealing
with the nuclear degrees of freedom, both classically and
quantum-mechanically.
  Large-scale first-principles condensed-matter and molecular 
simulations are extremely versatile, but most of their applicability 
demands an efficient treatment of both electrons and nuclei,
which will benefit from ($i$) improved robust communication between
nuclear-dynamics drivers and electronic-structure engines on
varied platforms, and
($ii$) hierarchical parallelisation of the integrated code,
to allow very large scale simulations on massively parallel
computers. 
  Library solutions for both problems and, especially, their 
integration, represents a promising direction for the community, 
building on initiatives such as i-PI.\cite{KAPIL2019214}
  It is a line of work that would involve the core of the CECAM
community, not only the electronic side, representing 
a great opportunity for the future of condensed-matter
and molecular simulations.

\section{Conclusions}

  The electronic structure library project presented here
is an initiative to stimulate, coordinate and amplify the efforts 
in library sharing already started within the electronic structure
community.
  It was initiated by CECAM, which continues its support together 
with the E-CAM European Centre of Excellence, spearheading a 
push within the community for a better model of electronic structure 
software development which, it is hoped, will enhance dynamism, versatility, 
maintainability and optimisation of electronic structure codes.
  It will rationalise coding effort by avoiding useless repetition,
and by separating different types of coding task to be carried out by people
with suitable profiles and backgrounds, distinguishing between 
computational scientists and computer scientists or software
engineers.
  We believe that it will allow the re-engineering efforts needed for 
deployment of electronic codes on novel computer architectures to be 
carried out more efficiently, widely, and by professionals close to 
hardware companies and HPC centres.

  Importantly, it is a community effort, pushed by people involved 
in the development of very prominent and popular electronic structure
codes, representing a wide spectrum of the community.
  Most of the library packages presented in this paper
were extracted from those codes, and many of these are currently being
used by codes other than their parent codes.
  There has been an emphasis on library packages for highly-parallel
heavy-duty tasks, the sharing of which is more challenging, but 
very important for the ambitions of the \esl.

  In addition to extracting, generating, and documenting
the library packages and adapting their APIs for general use,
part of the \esl effort is dedicated to facing the new 
challenges arising with the model.
  Most prominently, the integration of units with different 
data structures and parallelisation, and the bundling of the
set of packages in the \esl library for consistent and 
automatic building and compiling.

  Finally, as a community effort, the \esl community welcomes new 
additions to the \esl, and, of course, the use of the \esl or its 
components by any electronic structure programmer, or indeed any 
other community, as well as user feedback.

\section*{Authors contributions}

  All authors have contributed to this paper by coding and organizing
the coding events. 
  Micael J.~T.~Oliveira, Nick Papior, Yann Pouillon and Volker Blum have 
contributed to the \esl by coordinating the project at coding events and
in between them;
  Micael J.~T.~Oliveira, Nick Papior, and Yann Pouillon have maintained 
the software infrastructure.
  Most authors have contributed to the writing of the paper, either of
particular sections or by revising it. 
  Micael~J.~T. Oliveira and Emilio Artacho have coordinated the writing.

\begin{acknowledgments}
  The authors would like to thank CECAM for launching and pushing 
the \esl, as well as hosting part of its infrastructure, and partly
funding the extended workshops where most of the coding was done, both
in the Lausanne headquarters as in the Dublin, Trieste and Zaragoza nodes.  
Within CECAM, particular thanks to Sara Bonella, Bogdan Nichita, and
Ignacio Pagonabarraga.
  The authors also acknowledge all the people that have supported 
and contributed to the \esl in different ways, including  
Luis Agapito,
Xavier Andrade,
Balint Aradi,
Emanuele Bosoni,
Lori A. Burns,
Christian Carbogno,
Ivan Carnimeo,
Abel Carreras Conill,
Alberto Castro,
Michele Ceriotti,
Anoop Chandran,
Wibe de Jong,
Pietro Delugas,
Thierry Deutsch,
Hubert Ebert,
Aleksandr Fonari,
Luca Ghiringhelli,
Paolo Giannozzi,
Matteo Giantomassi,
Judit Gimenez,
Ivan Girotto,
Xavier Gonze,
Benjamin Hourahine,
J\"urg Hutter,
Thomas Keal,
Jan Kloppenburg,
Hyungjun Lee,
Liang Liang,
Lin Lin,
Jianfeng Lu,
Nicola Marzari,
Donal MacKernan,
Layla Martin-Samos,
Paolo Medeiros,
Fawzi Mohamed,
Jens Jørgen Mortensen,
Sebastian Ohlmann,
David O'Regan,
Charles Patterson,
Etienne Pl\'esiat,
Markus Rampp,
Laura Ratcliff,
Stefano Sanvito,
Paul Saxe,
Matthias Scheffler,
Didier Sebilleau,
Søren Smidstrup,
James Spencer,
Atsushi Togo,
Joost Vandevondele,
Matthieu Verstraete,
and Brian Wylie.

The authors would also like to thank the Psi-k network for having partially funded several of the \esl workshops.
  Alan O'Cais, Emilio Artacho, David L\'opez-Durán, Stefano de Gironcoli,
Emine K\"{u}\c{c}\"{u}kbenli, Arash Mostofi, and Mike Payne have received funding from the 
European Union's Horizon 2020 research and innovation program under the grant 
agreement No. 676531 (Centre of Excellence project E-CAM).
  The same project has partly funded the extended software development
workshops in which most of the \esl coding effort has happened.
  Alberto Garc\'{\i}a, Stephan Mohr, and Emilio Artacho acknowledge support
from the European Union’s Horizon 2020 research and innovation program 
under the grant agreement No. 824143 (Centre of Excellence project MaX).
  Miguel A.L. Marques acknowledges partial support from the DFG 
through the project MA-6786/1.
  Daniel G.A. Smith was supported by U. S. National Science Foundation 
(NSF) grant ACI-1547580.
  Mike Payne acknowledges support from EPSRC under grant EP/P034616/1.
  Arash Mostofi acknowledges support from the Thomas Young Centre under 
grant TYC-101, the Wannier Developers Group and all of the authors and 
contributors of the \wannier\ code (see Ref.~\onlinecite{wannier-github} 
for a complete list).
Alin M. Elena acknowledges support by CoSeC, the Computational Science Centre for Research Communities, through CCP5: The Computer  Simulation of Condensed Phases, EPSRC grants  EP/M022617/1 and EP/P022308/1.
  Alberto Garc\'{\i}a and José M. Soler acknowledge grant 
PGC2018-096955-B-C42 from Spain's Ministry of Science.
  Emilio Artacho, Alberto Garc\'{\i}a and José M. Soler acknowledge grant 
FIS2015-64886-C5 from Spain's Ministry of Science.
  Yann Pouillon, David L\'opez-Dur\'an and Emilio Artacho acknowledge 
support from grant RTC-2016-5681-7 from Spanish MINECO and EU Structural
Investment Funds.
  Martin L{\"u}ders acknowledges support from EPRSC under grant EP/M022668/1.
Martin L{\"u}ders, Micael J.~T.~Oliveira, and Yann Pouillon acknowledge 
support from the EU COST action MP1306.
  Jan Minar was supported by the European Regional Development 
Fund (ERDF), project CEDAMNF, reg. no.
CZ.02.1.01/0.0/0.0/15-003/0000358.
Victor Wen-zhe Yu, William Paul Huhn, Yingzhou Li, and Volker Blum 
acknowledge support from the National Science Foundation under award number 
ACI-1450280 (the ELSI project).
  Victor Wen-zhe Yu furthermore acknowledges a MolSSI fellowship (NSF award ACI-1547580). 
  Simune Atomistics S.L. is thanked for their allowing Ask Larsen and
Yann Pouillon to contribute to \esl, as is Synopsys Inc. for 
Fabiano Corsetti's partial availability.
\end{acknowledgments}

\section*{Data Availability Statement}

Data sharing not applicable --- no new data generated. 
Repositories for all \esl software packages mentioned in this work
have been properly cited, and are publicly available.

\appendix

\section{Community organization and Steering Structure of the \esl}
\label{sec:steering}

Formally, the \esl project was kick-started at a workshop organized at
CECAM-HQ by Emilio Artacho, Mike Payne, and Dominic Tildesley. Hosting around 20
participants, the workshop was held during the summer of 2014 over a period of
six weeks. After extensive discussions, the objectives and scope of the library
were agreed and the basic infrastructure was put in place. At this point, the
key element was the \esl wiki containing information about existing libraries
and modules. Also at this time, a governance structure was put into place consisting in a Curating Team (CT) and a
Scientific Advisory Board.

Since then, more workshops have been organized, roughly one per year, where
the \esl has been changed, improved, and expanded. It has evolved from a repository
of information about software libraries and tools in the domain of electronic
structure to a curated bundle of tightly integrated software libraries. As the
project evolved and mutated, its governance adapted to better
serve its objectives. In 2019, the Advisory Board was replaced by a Steering Committee (SC).
The SC proposes and defines the guidelines that the CT should follow. There are quarterly meetings which are open to the public and which focus on at least 3 tasks: 
\textit{i})  deciding which new libraries should be added to the \eslbundle, 
\textit{ii}) proposing which versions of existing \esl software should be shipped and 
\textit{iii}) discussions of topics for coming workshops.
The SC aims to include as many developers of software included in the \eslbundle and from codes using it as possible. It currently has 12 members and all CT members are part of the SC as well.
More recently, the curating team has been expanded from 3 to 6 members.
The CT manages everyday activities within the \esl by holding monthly meetings, creating proposal drafts and communicating with code developers and the SC.
Each member of the CT is tasked with supervising one specific aspect of the \esl. These include, amongst others, bundle maintenance, 
organization of the \esl workshop, the \esl website, and \esl documentation. 
Note that there are no competing interests between CT and SC. 
The \esl initiative aims to hold at least one workshop a year. These have a duration of 14 days of which 2 days are for discussions and the remaining 12 days of are for hands-on development activities. 
The focus of the workshops shifts each year with the topic decided by the SC.

\section{Sustainability and software engineering of the \esl demonstrator} 

Continuous Integration (CI) is a software engineering practice which allows code integration from multiple contributors automatically into the main repository of a project. The process is enabled by a set of tools and stages that assert the correctness of the code at each change. We strongly believe that CI is a critical requirement for any scientific software project, in order to maintain a sufficient level of quality over time.

CI is used within the ESL in a systematic way for the development of the \eslbundle{} and the \esldemo{}, as well as to check that the \esldemo{} can keep relying on new versions of the \eslbundle{}. The \esldemo{} is a basic example of ESC code built exclusively with ESL components to explain to developers how to use them in their own codes (see section \ref{sec:esl-demo}). As such, it has to be permanently kept in a working state. Some of the individual components of the ESL also benefit from CI, upon choice of their respective developers.

\begin{figure}
  \includegraphics[width=6cm, keepaspectratio]{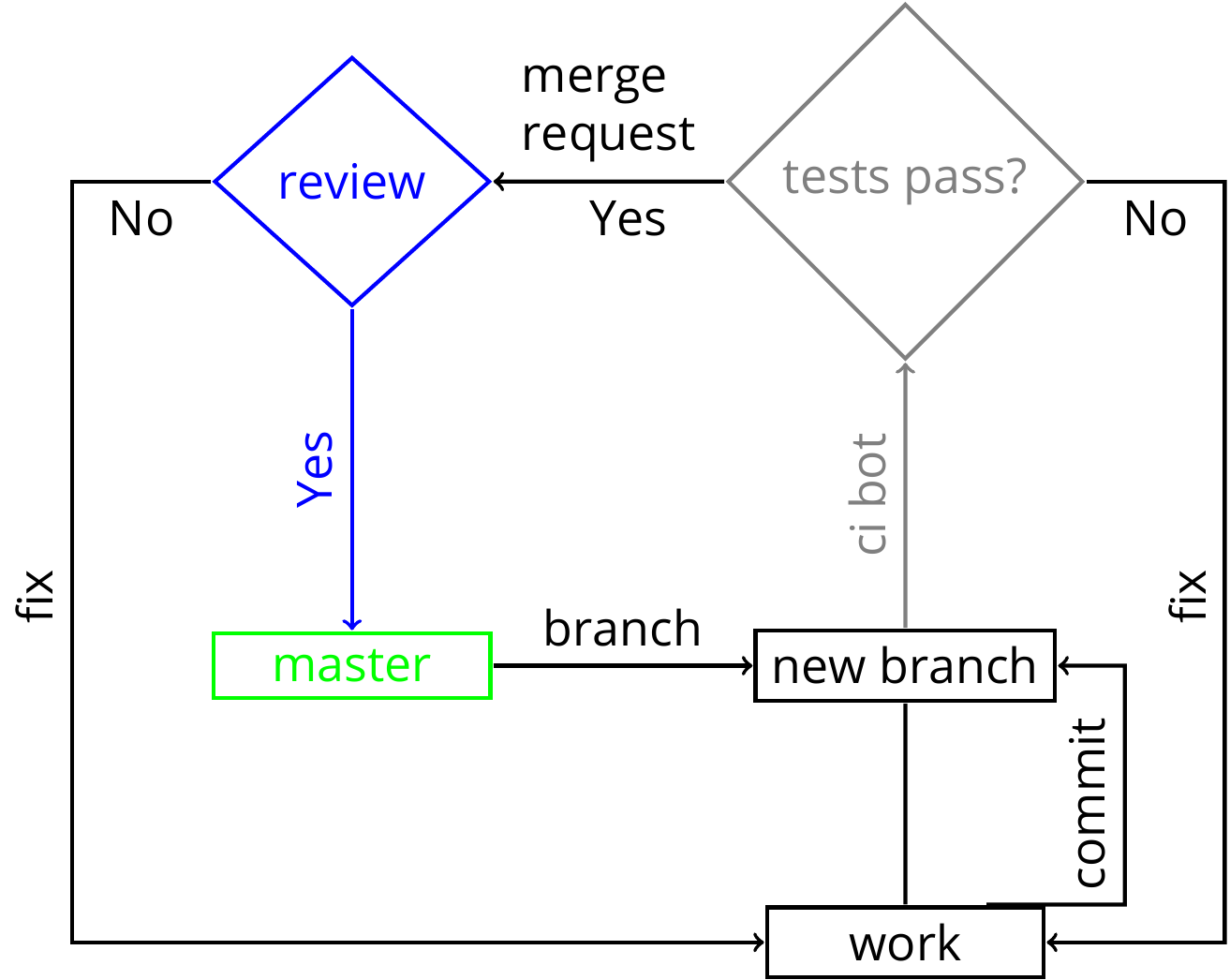}
  \caption{Continuous integration workflow integrated with git.}
  \label{fig:ci-workflow}
\end{figure}

As an example, the \esldemo relies on a series of widespread tools and technologies: Gitlab~CI\cite{gitlab-ci}, CTest from CMake\cite{ctest}, YAML\cite{yaml} and Docker/Docker Hub\cite{docker}. All these ingredients are glued together to provide development workflows implemented in the ESL (see Fig~\ref{fig:ci-workflow}). After each commit, the CI infrastructure automatically checks that the code successfully builds and the corresponding tests pass. 

Docker is one of the tools designed to help with running and deployment of applications by using container technology. It is relatively straightforward to use and we deploy the entire \eslbundle{} on it. We offer three Linux flavours for our Docker images: Ubuntu, Fedora and OpenSUSE. These Docker images are the ones we use in the build and testing stage. The images are public and distributed via Docker Hub.

The EasyBuild framework is of great help in this context. It has support for generating Singularity and Docker container recipes which will use EasyBuild to build and install reference software stacks. The latter will then be used within the CI infrastructure of the ESL, which mostly uses container-based runners in the cloud.

Gitlab CI is the integrated Gitlab tool for continuous integration, continuous delivery and deployment, and is highly configurable. At the moment we are using it only for CI.

CTest is a testing tool distributed as part of CMake. It integrates seamlessly with CMake, which is our build system of choice and one can easily use it to run unit tests or regression tests. We choose to use the latter due to the lack of any maintained unit testing framework for Fortran. A test is deemed passed or failed based on matching a regular expression at the end of execution. We also defined a target that monitors the code coverage of our tests. In order to help with the automation of testing, the output of the \esldemo is YAML-compliant, helping us to easily check the output. For checking output we rely on a simplified version of \siesta's YAML output testing. 

In the \esl workflow (see Fig.~\ref{fig:ci-workflow}), the user starts from a validated version of the \textit{master} repository by branching their own branch, a \textit{user branch}. Once the work envisaged is done and the user commits the changes to the branch, the Gitlab CI automatically runs the designed tests. If the tests fail, the user corrects the errors and commits again. Once the tests pass, the code is ready for a \textit{merge request}. A merge request is issued by the user for inclusion in the master branch of the project. A \textit{peer review process} then kicks in. Two reviewers have to agree for the code to be included in the master branch. If issues are found the code is returned to the user to fix the issues. If both reviewers agree then the code is integrated.


%

\end{document}